\documentclass[aps,twocolumn,superscriptaddress,floatfix,nofootinbib]{revtex4-2}

\usepackage{amsmath}
\usepackage{siunitx}
\usepackage[inline]{enumitem}

\usepackage{xcolor}  % For corrections
\usepackage{graphicx}
\usepackage[section]{placeins}
\usepackage{amsfonts}
\usepackage{mathtools}
\usepackage{booktabs}
\usepackage{braket}
\usepackage{MnSymbol}
\usepackage{tikz}
\usepackage{cancel}
\usepackage{setspace}
\usepackage{upgreek}
\usepackage{amsthm}
\usepackage{bbm}
\makeatletter
\newtheorem*{rep@theorem}{\rep@title}
\newcommand{\newreptheorem}[2]{%
\newenvironment{rep#1}[1]{%
 \def\rep@title{#2 \ref*{##1}, repeated}%
 \begin{rep@theorem}}%
 {\end{rep@theorem}}}
\makeatother

\newcommand{\Id}{\mathbbm{1}}

\newtheorem{lemma}{Lemma}
\newreptheorem{lemma}{Lemma}
\newtheorem{theorem}{Theorem}

\newreptheorem{proposition}{Proposition}
\newreptheorem{corollary}{Corollary}

\newcommand{\abs}[1]{\left | {#1}\right |}
\newcommand{\norm}[1]{\left \Vert {#1}\right \Vert}
\newcommand{\smallabs}[1]{ | {#1} |}
\newcommand{\smallnorm}[1]{ \Vert {#1} \Vert}

\begin{document}
\title{Efficient representations by PEPS:  critical free-fermion and chiral topological states}
\author{Ignacio Cirac}
\email{ignacio.cirac@mpq.mpg.de}
\author{Rahul Trivedi}
\email{rahul.trivedi@mpq.mpg.de}
\affiliation{Max-Planck-Institut f\"ur Quantenoptik, Hans-Kopfermann-Stra\ss e 1, D-85748 Garching, Germany}
\affiliation{Munich Center for Quantum Science and Technology (MCQST), Schellingstra\ss e 4, D-80799 M\"unchen, Germany}
\begin{abstract}
We construct projected entangled-pair state (PEPS) approximations to various many-body states in 2D lattices with global error at most $\epsilon$ and bond dimension polynomial in the system size and $1/\epsilon$. They include critical free-fermion models in two dimensions, as well as chiral topological states. For the two-dimensional nearest-neighbor tight-binding model at half filling, our construction combines Gaussian wavelet transformations with matrix product state approximations of the one-dimensional chains, as well as selected fermionic swap operations. For arbitrary uniformly gapped, finite-range quadratic Hamiltonians, we give a different construction based on repeated one-particle filtering, which yields Gaussian PEPS in any  spatial dimension. Those include, for instance, $p+ip$ models or a Chern-insulator in a square lattice, for which we give an explicit construction. Finally, we prove that Gutzwiller projections of a fixed number of identical Gaussian PEPS approximations to free-fermion states at arbitrary filling do not amplify considerably the approximation error. This provides a polynomial-bond PEPS approximations to the resulting states, including familiar lattice chiral-spin-liquid states.
\end{abstract}
\maketitle
\raggedbottom

% ======================================================
\section{Introduction}
\label{sec:introduction}

Tensor networks provide succinct descriptions of quantum many-body states in terms of local tensors~\cite{orus2014review,cirac2021rmp}. On the one hand, they have become indispensable tools in computational many-body physics. Indeed, matrix product states (MPS) underlie the density-matrix renormalization group (DMRG) and related one-dimensional methods~\cite{white1992dmrg,schollwoeck2011}, while projected entangled-pair states (PEPS) extend these approaches to higher dimensions and are used to approximate ground and thermal states~\cite{verstraete2004PEPS,verstraete2006criticality}. On the other hand, the tensors themselves encode physical properties of the states. Their symmetries and algebraic structure provide ways of characterizing quantum phases, most notably topologically ordered phases in two dimensions~\cite{levinwen2005stringnet,aguadovidal2008topological,buerschaper2009stringnet,schuch2010topology,bultinck2017anyons} and symmetry-protected topological phases~\cite{chenguwen2011spt,pollmann2012spt,schuch2011phases,chen2012spt}. Thus, tensor networks are not merely computational ansatzes, but have also become a useful framework for understanding quantum many-body physics.

A central question in the context of tensor networks is whether physically relevant states admit efficient approximations. Here, efficiency means that, on a lattice of linear size $n$, a state can be approximated to arbitrary prescribed error $\epsilon$ using a bond dimension $D$ that scales polynomially with $n$ and $1/\epsilon$. Such an approximation provides a compact classical description in terms of a polynomial number of parameters, rather than the exponentially many amplitudes required to specify a many-body wavefunction. This exponential growth is a central obstacle to the classical simulation of quantum systems and was one of the motivations behind Feynman's proposal to simulate quantum physics using quantum computers~\cite{feynman1982simulating}. For Gibbs states of local Hamiltonians at any fixed nonzero temperature, polynomial PEPS approximations were established in Ref.~\cite{molnar2015gibbs}, building on the earlier work of Hastings~\cite{hastings2006local}. At zero temperature, ground states of gapped one-dimensional local Hamiltonians admit efficient MPS approximations~\cite{hastings2007area}, and subsequent work strengthened the underlying area-law and algorithmic results~\cite{arad2017rg}. In two dimensions, ground-state approximation by PEPS required additional conditions or addressed more restricted questions~\cite{hastings2006local,schuch2007complexity,molnar2015gibbs,anshuarad2016local,openai2026peps}. These developments are closely linked to entanglement area laws (see, e.g., Ref.~\cite{eisert2010review} for a review). An area law by itself, however, does not generally guarantee an efficient PEPS approximation in dimensions higher than one.

The situation is less clear for critical systems, which are central to the description of phase transitions, as well as for states displaying chiral topological order. In one dimension, an upper bound of order $\log n$ on the R\'enyi entropy $S_{1/2}$ across every cut suffices for an MPS approximation with $D=\operatorname{poly}(n,1/\epsilon)$~\cite{verstraetecirac2006,schuch2008entropy}. Critical free-fermion chains satisfy such a bound and can therefore be efficiently approximated by general MPS. In two dimensions, many familiar critical points retain a leading area-law contribution, with the form $S(L)=aL+b\log L+O(1)$ for polygonal regions when logarithmic corner terms are present~\cite{fradkinmoore2006corner,casinihuerta2007corners,stoudenmire2014corners}. The logarithmic term is subleading, and the presence or absence of such a term depends on the critical theory and the shape of the region. In contrast, fermionic models (like the tight-binding model) possess an extended Fermi surface, leading to the stronger violation $S(L)=\Theta(L\log L)$~\cite{wolf2006violation,gioev2006widom,barthel2006entanglement} and therefore they appear to be the most difficult to describe in terms of tensor networks. Indeed, for such a two-dimensional critical system, efficient approximability has remained as a relevant open question. Note that numerical Gaussian tensor-network descriptions of Fermi surfaces~\cite{mortier2022surfaces} do not by themselves settle this issue at arbitrarily small global error.

Gapped chiral states present a different challenge for tensor-network descriptions. In the Gaussian setting, exact translationally invariant PEPS of fixed finite bond dimension with a nonzero Chern number cannot have a strictly local, gapped quadratic parent Hamiltonian; their local frustration-free quadratic parents must therefore be gapless~\cite{wahl2013chiral,dubail2015chiral}. Moreover, a nonzero Chern number obstructs an orthonormal basis of exponentially localized Wannier functions spanning the occupied bands~\cite{brouder2007wannier}. More generally, these results have raised the question of how faithfully finite-bond-dimension PEPS can represent gapped chiral topological order, such as lattice analogues of Laughlin states. At the same time, explicit chiral PEPS constructions and numerical studies give substantial evidence that PEPS can capture characteristic properties of such phases, including chiral edge spectra and competitive variational energies~\cite{wahl2013chiral,wahl2014symmetries,yang2015chiral,poilblanc2015chiral,poilblanc2017investigation,chen2018nonabelian}. Thus, while exact finite-bond-dimension representations are subject to important obstructions, PEPS remain promising variational descriptions of gapped chiral topological phases. However, these examples do not establish that such descriptions are efficient: the bond dimension required to achieve a prescribed global approximation error $\epsilon$ could still grow exponentially with the system size or with $1/\epsilon$.

In this work we establish efficient PEPS approximations for states of both kinds in models. In both cases, the global error is at most $\epsilon$ and the bond dimension is $D=\operatorname{poly}(n,1/\epsilon)$. We specifically consider the two dimensional gappless tight-binding model at half filling, uniformly gapped Gaussian models (including chiral ones, like Chern insulators or $p+ip$), as well as the ones obtained by Gutzwiller projecting several copies of those states. 

For the tight-binding model, our construction begins with an auxiliary free-fermion model whose occupied orbitals factorize into one-dimensional components. Using the connection between wavelet transformations and entanglement renormalization~\cite{evenbly2016wavelets,haegeman2018wavelets}, we separate at each scale a smaller copy of the two-dimensional problem, one-dimensional critical chains, and local dimers. We approximate the Gaussian wavelet transformations by exactly unitary matrix product operators (MPUs), and use fermionic SWAPs to arrange successive scales on different rows and columns. The critical chains are replaced by MPS approximations, and reversing the transformation yields a PEPS with a polynomially bounded bond dimension and controlled error. 

An important feature of the tight-binding construction is that the resulting PEPS need not be Gaussian, even though the target state is a free-fermion Gaussian state. Indeed, it was proved that Gaussian fermionic MPS cannot approximate the critical one-dimensional hopping ground state with polynomial bond dimension, whereas unrestricted fermionic MPS can~\cite{francorubio2022gaussian}. A related obstruction arises for Gaussian PEPS in two dimensions. Our construction avoids the Gaussian restriction precisely when the released one-dimensional chains are replaced by general MPS. In fact, all remaining transformations, including the approximate MPUs and the fermionic SWAPs, are Gaussian.

For gapped Gaussian models, we use a different construction. We choose an occupation product state whose overlap matrix with the occupied subspace has a polynomially bounded inverse, and apply a fixed-range one-particle filter repeatedly. The error decreases exponentially with the number of steps, while each induced many-body map has a constant-bond Gaussian PEPO representation. This gives a Gaussian PEPS approximation with polynomial bond dimension in any fixed spatial dimension. A local doubling transformation and a controlled projection onto one copy extend the argument to Hamiltonians with pairing, including gapped $p+ip$ models~\cite{readgreen2000}. We also provide a more direct construction for the square-lattice Chern insulator~\cite{hatsugai1990chern,nielsen2013local}.

Finally, we use these results to describe chiral topological states obtained by Gutzwiller projecting several copies of Chern-band states~\cite{kalmeyer1987rvb,jain1989parton,wen1991nonabelian}. For the equal-occupancy projection of a fixed number of identical free-fermion states, we show that the normalized projected state again has a PEPS approximation with bond dimension $D=\operatorname{poly}(n,1/\epsilon)$. The proof introduces a stability estimate: particle-number-conserving Gaussian states can be projected into occupied and unoccupied modes, and their normalized wavefunctions differ by at most a polynomial amplification of the input error. The argument works at arbitrary filling and does not divide by the exponentially small norm of the unnormalized projected state. For two copies of the square-lattice Chern insulator state it give the chiral states analyzed in ~\cite{zhang2011topological}. Higher Chern number generalizations are studied in Ref.~\cite{zhang2013nonabelian}.

% ======================================================
\section{Models}
\label{sec:freeFermionModels}

We consider several models on lattices, beginning with the one- and two-dimensional fermionic tight-binding problems, and then introducing gapped fermioninc Gaussian Hamiltonians and their Gutzwiller-projected states. The tight-binding models have periodic boundary conditions, whereas our explicit Chern-insulator example has antiperiodic boundary conditions. We name the lattice sites along one direction by $x\in\mathcal X=\{0,\ldots,n-1\}$; for the tight-binding construction $n=2^\ell$, while the Chern example allows any even $n\geq4$. For periodic boundary conditions, we use momenta $k\in\mathcal K=\{-\pi+2\pi z/n:z\in\mathcal X\}$. Sites and momenta are understood modulo $n$ and $2\pi$, respectively. For the general Gaussian and projected models, $N$ denotes the total number of fermionic modes, with a fixed number per site. Equalities between many-body states are understood up to consistently chosen overall phases.

% ------------------------------------------------------
\subsection{Tight-binding model in one dimension}
\label{sec:tightBinding1D}

We consider first a ring with $n$ sites and denote by $a_x$ the annihilation operator at lattice site $x=0,\ldots,n-1$. The tight-binding Hamiltonian is
\begin{equation}
    \label{eq:hamiltonian1D}
    H_1 = - \sum_{x\in \mathcal X} (a^\dagger_x a_{x+1} + H.c.),
\end{equation}
The total number of fermions is conserved. Since we are looking at the ground state and there is particle-hole symmetry, the number of fermions is half of the number of sites, i.e. we work at half filling.

We will also consider another related model which is better suited for the wavelet description. It has two modes $(c_x,d_x)$ per unit cell and is defined as
\begin{equation}
    \label{eq:auxiliaryHamiltonian1D}
    \tilde H_1 = - \sum_{x\in \mathcal X} \left[c^\dagger_{x} (d_{x} - d_{x-1}) +  H.c.\right],
\end{equation}
It is related to the previous one with double the lattice sites just by taking $c_x=(-1)^x a_{2x},d_x=(-1)^x a_{2x+1}$.

Sometimes we will use momentum representations
\begin{equation}
 c_x=\frac1{\sqrt n}\sum_{k\in\mathcal K}e^{i kx }\hat c_k,
 \qquad
 d_x=\frac1{\sqrt n}\sum_{k\in\mathcal K}e^{i kx}\hat d_k.
 \label{eq:fourierTransform1D}
\end{equation}
We can thus write
\begin{equation}
 \tilde H_1=-\sum_{k\in\mathcal K}
 \begin{pmatrix}\hat c_k^\dagger&\hat d_k^\dagger\end{pmatrix}
 \begin{pmatrix}0&f(k)\\\overline{f(k)}&0\end{pmatrix}
 \begin{pmatrix}\hat c_k\\\hat d_k\end{pmatrix},
 \label{eq:momentumHamiltonian1D}
\end{equation}
where $f(k)=2i\,e^{-i k/2}\sin(k/2)$.

The ground state can be written as
\begin{equation}
    \label{eq:groundState1D}
    \ket{\omega} =\prod_{k\in\mathcal K}\beta_k^\dagger\ket0,\qquad \beta_k^\dagger=  \frac{\gamma(k)\hat c_k^\dagger+\hat d_k^\dagger}{\sqrt2},
\end{equation}
where $\ket0$ is the state containing no fermions,
\begin{equation}
 \label{eq:phaseGamma}
 \gamma(k)=
 \begin{cases}
 \displaystyle\frac{f(k)}{|f(k)|}  =i\,{\rm sgn}(k)e^{-i k/2},&k\ne0,\\ \quad
 1,&k=0.
 \end{cases}
\end{equation}
and we have deliberately chosen one of the two degenerate ground states.

It will be convenient to use a compact notation for the operators. We define column vectors $\boldsymbol c$ and $\boldsymbol d$ of annihilation operators, with adjoint row vectors
\begin{equation}
    \label{eq:display001}
   \boldsymbol c^\dagger = (c_0^\dagger,\ldots,c_{n-1}^\dagger), \qquad
   \boldsymbol d^\dagger = (d_0^\dagger,\ldots,d_{n-1}^\dagger)
\end{equation}
We can also define $\boldsymbol \beta$ the vector that contains all the operators $\beta_k$ sorted according to the values of $k$.  We can write Eq.~\eqref{eq:groundState1D} as
\begin{equation}
    \label{eq:display002}
    \boldsymbol \beta^\dagger = \frac{ \boldsymbol c^\dagger F \Gamma +\boldsymbol d^\dagger F}{\sqrt2},
\end{equation}
where $\Gamma$ is a diagonal matrix with entries $\gamma(k)$ and $F$ is the Fourier matrix  with elements $(F)_{x,k}=e^{ikx}/\sqrt{n}$.

Note that in the ground state all the $\beta$ modes are occupied, and the rest are empty. We could have chosen any linear combinations of those modes to define the ground state. This means that if we define new modes $V\boldsymbol \beta$, where $V$ is unitary, we can re-express the ground state in terms of those modes. A particularly relevant choice is $V=F$ which yields
\begin{equation}
    \label{eq:display003}
    \boldsymbol \eta^\dagger =  \frac{ \boldsymbol c^\dagger G +\boldsymbol d^\dagger}{\sqrt2}
\end{equation}
where we have used $FF^\dagger=\Id$ and defined
\begin{equation}
    \label{eq:matrixG}
    G = F \Gamma F^\dagger, \quad (G)_{x,y}= \frac{1}{n}\sum_{k\in\mathcal K}\gamma(k)e^{ik(x-y)},
\end{equation}
Note that $\boldsymbol \eta$ is the (inverse) Fourier transform of $\boldsymbol \beta$, and can be viewed as a vector of components $\eta_x$, with $x\in \mathcal X$. Thus, we can also write the ground state as
\begin{equation}
    \label{eq:realSpaceGroundState1D}
    \ket{\omega} =\prod_{x\in\mathcal X}\eta_x^\dagger\ket0.
 \end{equation}

% ------------------------------------------------------
\subsection{Tight-binding model in two dimensions}\label{sec:tightBinding2D}

In an $n\times n$ lattice with periodic boundary conditions, we have
\begin{equation}
    \label{eq:hamiltonian2D}
    H_2 = - \sum_{x,y=0}^{n-1} \left[a^\dagger_{x,y} (a_{x+1,y} + a_{x,y+1}) +  H.c.\right],
\end{equation}
where the ground state is again reached at half filling.

Following Ref.~\cite{haegeman2018wavelets}, we consider another related model which will be more amenable to the wavelet method. It has two modes $(c,d)$ per unit cell and is defined as
\begin{equation}
    \label{eq:auxiliaryHamiltonian2D}
    \begin{aligned}
        \tilde H_2=-\sum_{x,y=0}^{n-1}\bigl[&c^\dagger_{x,y}(d_{x,y}-d_{x-1,y}\\
        &-d_{x,y-1}+d_{x-1,y-1})+H.c.\bigr],
    \end{aligned}
\end{equation}
We can write it in momentum representation as
\begin{equation}
    \label{eq:display004}
    \tilde H_2 =-\sum_{\mathbf k} \begin{pmatrix} \hat c_{\mathbf k}^\dagger & \hat d_{\mathbf k}^\dagger     \end{pmatrix}
    \begin{pmatrix} 0 & f(k_x)f(k_y)\\ \overline{f(k_x)f(k_y)} & 0 \end{pmatrix}
    \begin{pmatrix} \hat c_{\mathbf k}\\ \hat d_{\mathbf k} \end{pmatrix}.
\end{equation}
where $\mathbf k=(k_x,k_y)$ and $k_x,k_y\in \mathcal K$.

The ground state can be written as in Eq.~\eqref{eq:groundState1D}, with creation operators
\begin{equation}
    \label{eq:display005}
    \beta_{\mathbf k}^\dagger= \frac{\gamma(k_x)\gamma(k_y) \hat c_{\mathbf k}^\dagger+  \hat d_{\mathbf k}^\dagger}{\sqrt2},
\end{equation}
and $\gamma(k)$ is defined in Eq.~\eqref{eq:phaseGamma}. As in one dimension, we have chosen one of the degenerate ground states.

As before, we now introduce the column vectors $\boldsymbol c$, $\boldsymbol d$ containing all the corresponding operators, ordered by the first coordinate and then by the second. We similarly define $\boldsymbol\beta$ with the momenta ordered in the same way. As in one dimension, we can make unitary transformations among the occupied modes without changing the ground state. In particular, defining
$\boldsymbol \eta = (F\otimes F)\boldsymbol\beta$, we obtain
\begin{equation}
    \label{eq:display006}
    \boldsymbol\eta^\dagger= \frac{ \boldsymbol c^\dagger(G\otimes G) + \boldsymbol d^\dagger }{\sqrt2}
\end{equation}
where $G$ is defined in Eq.~\eqref{eq:matrixG}. Note the tensor product structure corresponds to the
decomposition into rows and columns. Thus, the ground state can also be expressed as
\begin{equation}
    \label{eq:realSpaceGroundState2D}
    \ket{\Omega} =\prod_{x,y\in\mathcal X}\eta_{x,y}^\dagger\ket0.
 \end{equation}
Note that we use capital $\Omega$ to denote the ground state in 2D in order to distinguish it from
the one in 1D.

% ------------------------------------------------------
\subsection{Gapped Gaussian models}
\label{sec:gappedGaussianModels}
\label{sec:chernModel}

We now consider gapped Gaussian Hamiltonians. Let $a_1,\ldots,a_N$ denote fermionic annihilation operators, with mode $i$ at lattice position $x_i$. There is a fixed number of modes per site. We consider
\begin{equation}
    \label{eq:generalGaussianHamiltonian}
    H=\sum_{i,j}J_{ij}a_i^\dagger a_j
    +i\sum_{i,j}\left(G_{ij}a_i a_j-\mathrm{H.c.}\right),
\end{equation}
where $J^\dagger=J$ and $G^T=-G$ and both matrices may have complex entries. nb vWe assume finite-range interactions, namely $J_{ij}=G_{ij}=0$ whenever $d(x_i,x_j)>R$, with $R$ independent of the system size. To fix a normalization, we also assume that
\begin{equation}
    \label{eq:normalization}
    \max_i\sum_j|J_{ij}|<1,
    \qquad \max_i\sum_j|G_{ij}|<1.
\end{equation}

We will assume that the Hamiltonian $H$ has a unique ground-state and an energy gap. To formalize this assumption in terms of the coefficients $J_{i, j}$ and $G_{i, j}$, we express the Hamiltonian in Eq.~\eqref{eq:generalGaussianHamiltonian} in terms of the Majorana operators $\gamma_i^{0} = (a_i + a_i^\dagger)/\sqrt{2}, \gamma_i^1 = -i(a_i - a_i^\dagger)/\sqrt{2}$ to obtain
\begin{align}\label{eq:gaussian_H_majorana}
    H = \frac{1}{2} \text{Tr}(J)I+\sum_{i, j}\sum_{\alpha, \beta} h_{i, j}^{\alpha, \beta}\gamma_i^\alpha \gamma_j^\beta,
\end{align}
where
\begin{equation}\label{eq:h_in_terms_of_JG}
\begin{pmatrix}
h_{ij}^{0,0} & h_{ij}^{0,1} \\[2pt]
h_{ij}^{1,0} & h_{ij}^{1,1}
\end{pmatrix}
=
\frac{i}{2}
\begin{pmatrix}
\operatorname{Im} J_{ij} + 2\operatorname{Re} G_{ij}
&
\operatorname{Re} J_{ij} - 2\operatorname{Im} G_{ij}
\\[2pt]
-\operatorname{Re} J_{ij} - 2\operatorname{Im} G_{ij}
&
\operatorname{Im} J_{ij} - 2\operatorname{Re} G_{ij}
\end{pmatrix}.
\end{equation}
It can be seen that $h_{i, j}^{\alpha, \beta} \in i \mathbb{R}$ and $h_{i, j}^{\alpha, \beta} = -h_{j, i}^{\beta, \alpha}$ i.e., the matrix $h$ of the coefficients $h_{i, j}^{\alpha, \beta}$ with $(i, \alpha)$ being a row-index and $(j, \beta)$ being a column-index is a purely imaginary anti-symmetric matrix. Therefore, 
\begin{align}
\text{spec}(h) = \{\pm \varepsilon_1, \pm \varepsilon_2 \dots \pm \varepsilon_N\},
\end{align}
where $0\leq \varepsilon_1 \leq  \dots \leq  \varepsilon_N$ and by using an orthogonal transformation of the matrix $h$, we can rewrite $H$ as
\begin{align}
    H = \frac{1}{2}\text{Tr}(J) \Id + \sum_{i = 1}^N \varepsilon_i(2b_i^\dagger b_i - \Id),
\end{align}
where $b_i$ are fermionic annihilation operators $\{b_i, b_j^\dagger\} = \delta_{i, j}$ and $\{b_i, b_j\}=0$. In general $b_i$ will be related to $a_i$ by a Bogoliubov transformation i.e., $b_i = \sum_j u_{i, j} a_j + v_{i, j} a_j^\dagger$. In this representation, the vacuum of the $b-$modes is a ground-state $\ket{\Phi}$ of $H$ i.e., 
\begin{align}
    b_i \ket{\Phi} = 0 \ \forall i \in \{1,\dots , N\}.
\end{align}
The first excited state $b_1^\dagger \ket{0}$ will have energy $2\varepsilon_1$. Therefore, a necessary and sufficient condition for $H$ to have a unique ground-state and a many-body gap of $\Delta > 0$ is that 
\begin{align}\label{eq:gappedSpectrum}
    h^\dagger h \succeq \frac{\Delta^2}{4} \Id.
\end{align}
Note that, in the excitation number conserving case (i.e., $G = 0$), this condition reduces to $J^\dagger J \succeq \Delta^2  \Id$.

% First consider particle-number-conserving models, for which $G=0$. Since $\|J\|_{\mathrm{op}}\leq\max_i\sum_j|J_{ij}|<1$, the gap condition can be written as
% \begin{equation}
%     \label{eq:gappedSpectrum}
%     \operatorname{spec}(J)\subset(-1,-\Delta]\cup[\Delta,1),
%     \qquad 0<\Delta<1,
% \end{equation}
% where $\Delta$ is independent of $N$. The ground state is obtained by filling all negative-energy orbitals:
% \begin{equation}
%     \label{eq:gappedGroundState}
%     \ket{\Phi}=\prod_{i=1}^{r}a^\dagger(v^i)\ket0,
%     \qquad a^\dagger(u)=\sum_{j=1}^{N}u_j a_j^\dagger.
% \end{equation}
% Here $v^1,\ldots,v^r$ are orthonormal eigenvectors of $J$ with negative eigenvalues. Let $w^1,\ldots,w^{N-r}$ be orthonormal eigenvectors with positive eigenvalues, and denote by $V$ and $W$ the matrices containing these vectors as columns. Thus $V^\dagger V=1$, $W^\dagger W=1$, and $V^\dagger W=0$. 

We will be particularly interested in two-dimensional chiral models, such as Chern insulators and $p+ip$ superconductors~\cite{hatsugai1990chern,readgreen2000} both of which are gapped Gaussian models. As an explicit particle-number-conserving example, we consider the square-lattice $\pi$-flux model with diagonal hoppings~\cite{hatsugai1990chern,nielsen2013local}, whose occupied band is topologically nontrivial. In contrast to the tight-binding models, we impose antiperiodic boundary conditions in both directions, since this will simplify some of the statements. The Hamiltonian is
\begin{equation}
    \label{eq:chernHamiltonian}
    \begin{aligned}
        H_{\mathrm C}={}&\frac12\sum_{x,y\in\mathcal X}
        \Bigl[\bigl((-1)^y a_{x+1,y}^\dagger
        +a_{x,y+1}^\dagger\bigr)a_{x,y}+H.c.\Bigr]\\
        &+\frac14\sum_{x,y\in\mathcal X}
        \Bigl[i(-1)^y\bigl(a_{x+1,y+1}^\dagger
        -a_{x-1,y+1}^\dagger\bigr)a_{x,y}+H.c.\Bigr]\\
        ={}&\boldsymbol a^\dagger h\boldsymbol a .
    \end{aligned}
\end{equation}
The fixed energy convention in this example can be brought to Eq.~\eqref{eq:normalization} by dividing $H_{\mathrm C}$ by four, without changing its ground state. Here $\boldsymbol a$ is the column vector containing all $a_{x,y}$ in a fixed order, and $h$ is the corresponding one-particle matrix. As before, the coordinates are understood modulo $n$, and each crossing of a boundary, either horizontally or vertically, supplies a minus sign. A hopping crossing both boundaries has the product of the two signs. The nearest-neighbor hopping amplitudes give flux $\pi$ per plaquette; the imaginary next-nearest-neighbor (diagonal) hoppings open the gap. The simple two-band dispersion below is convenient for the polynomial approximation.

In Sec.~\ref{sec:gappedGaussianPEPS}, we construct a Gaussian PEPS approximation to the ground states of gapped Gaussian Fermionic models with polynomial bond dimension. We also provide a simpler and specific construction for the Chern insulator model described above, which maybe of independent interest.

% ------------------------------------------------------
\subsection{Gutzwiller-projected chiral topological states}
\label{sec:gutzwillerModels}

Many chiral topological states can be constructed by taking several copies of a fermionic Gaussian state and applying a local Gutzwiller projection~\cite{kalmeyer1987rvb,zhang2011topological,zhang2013nonabelian,tu2013projected}. This projection restricts the allowed occupations of the copies at each site and identifies the remaining states with physical spin or bosonic degrees of freedom. 

We will use explicitly the following family. Let $\ket{\Phi}$ have $r$ particles in $N$ modes, and let us consider a fixed number $m\geq2$ of identical copies. At each site, we define the local map
\begin{equation}
    \label{eq:gutzLocalMap}
    P_i=\ket0\bra{0\cdots0}+\ket1\bra{1\cdots1},
\end{equation}
and the projected state
\begin{equation}
    \label{eq:Psi0}
    \ket{\Psi_0}=\left(\prod_{i=1}^{N}P_i\right)\ket{\Phi}^{\otimes m},
    \qquad
    \ket{\Psi}=\frac{\ket{\Psi_0}}{\|\ket{\Psi_0}\|}.
\end{equation}
Thus all copies have the same occupation at each site. For $m=2$, this is the particle--hole-transformed form of the usual single-occupancy constraint: transforming the second copy converts $n_{i,1}+n_{i,2}=1$ into $n_{i,1}=n_{i,2}$. Away from half filling, the single-occupancy description uses $r$ particles in the first copy and $N-r$ particles in its particle--hole-transformed partner.

Writing $\ket{\Phi}=\sum_{|X|=r}\phi(X)\ket X$ in a fixed occupation basis, Eq.~\eqref{eq:Psi0} gives
\begin{equation}
    \label{eq:gutzPowerState}
    \ket{\Psi}=\frac{\sum_{|X|=r}\phi(X)^m\ket X}
    {\left(\sum_{|X|=r}|\phi(X)|^{2m}\right)^{1/2}}.
\end{equation}
The fermionic reordering between different copies contributes only a common sign at fixed particle number, which we omit here. For even $m$, it gives a spin or hard-core-boson state whereas for odd $m$ the composite particle is fermionic. Note that a fixed onsite phase associated with a particle--hole transformation can be included without affecting the approximation bounds.

For two filled $C=1$ Chern bands, this construction gives the determinant-square trial state for the Kalmeyer--Laughlin chiral spin liquid, with $\mathrm{SU}(2)_1$, or equivalently $\mathrm{U}(1)_2$, topological order~\cite{kalmeyer1987rvb,zhang2011topological}. In particular, the Chern model in Sec.~\ref{sec:gappedGaussianModels} is the square-lattice example used in this construction. More generally, powers of a filled $C=1$ band give lattice Laughlin trial states, while projected higher-Chern-number bands provide non-Abelian examples. For instance, two copies of a filled $C=2$ band give the determinant-square state described by $\mathrm{SU}(2)_2$ Chern--Simons theory coupled to fermions in Ref.~\cite{zhang2013nonabelian}.

There are related constructions using paired Gaussian states and other local constraints. Projecting two copies of a weak-pairing $p+ip$ state onto single occupancy gives a spin-$1/2$ chiral state with $\mathrm{SO}(2)_1\simeq\mathrm{U}(1)_4$ edge theory~\cite{tu2013projected,yang2015chiral}. Three flavors of weak-pairing $p+ip$ superconductors, with one fermion per site, give a spin-1 non-Abelian chiral spin liquid with $\mathrm{SO}(3)_1\simeq\mathrm{SU}(2)_2$ order~\cite{liu2018nonabelian}. Note that the approximation result proved below concerns the equal-occupancy Slater-determinant family in Eq.Eq.~\eqref{eq:gutzPowerState}.

% ======================================================
\section{1D tight-binding model: MPS and MPU approximations}
\label{sec:oneDimensionalApproximations}

In this Section, we consider tensor network approximations connected to the one-dimensional
tight-binding model described by the Hamiltonian in Eq.~\eqref{eq:auxiliaryHamiltonian1D}. First, we will consider the ground
state and construct an MPS approximation with precision $\epsilon$ and bond dimension
$D=\operatorname{poly}(n/\epsilon)$. Then, we will construct an MPU approximation to a wavelet transformation
$U$ which will be Gaussian and with a bond dimension scaling also as $D=$poly$(n,1/\epsilon)$.

% ------------------------------------------------------
\subsection{MPS description of the ground state}
\label{sec:groundStateMPS}

We consider the ground state $\ket{\omega}$ in Eq.~\eqref{eq:realSpaceGroundState1D} of the Hamiltonian
in Eq.~\eqref{eq:auxiliaryHamiltonian1D}. We will show that it admits an MPS approximation with polynomial bond
dimension. Note that the MPS we construct is not translationally invariant, although it is
straightforward to build one with such property at the expense of increasing $D$ by a factor of
$n$.

Following the standard sequential Schmidt-truncation construction of matrix product states~\cite{fannes1992,schollwoeck2011,verstraete2006criticality}, we proceed as follows. We choose a positive integer $D$, the maximum bond
dimension we will retain. We then construct the MPS by a left-to-right sequence of singular-value
decompositions: We first separate the first site from all remaining cells and retain the largest
$D$ terms in the corresponding Schmidt decomposition. Then we combine the retained left index with
the physical index of the second site and perform another singular-value decomposition, separating
these indices from the remaining sites and, again, retain at most $D$ terms. We continue in this
way until the last site. At each step, the previously constructed tensors remain fixed. Thus later
steps preserve the bond-dimension bounds on the earlier cuts. We keep the intermediate states
unnormalized and normalize only at the end.

To upper bound the error, we first define the Schmidt coefficients of the original exact state at
every cut between a left and a right subset of lattice sites. Let $A_a=\{0,\ldots,a-1\}$ consist
of the first $a$ sites. Then we write the Schmidt decomposition of $A_a$ with its complement as
\begin{equation}
    \begin{aligned}
        \ket{\omega} &= \sum_j\sqrt{\lambda_j}\,\ket{L_j}\ket{R_j},\\
        \lambda_1 &\geq\lambda_2\geq\cdots,
        \qquad \sum_j\lambda_j=1.
    \end{aligned}
    \label{eq:schmidtDecomposition}
\end{equation}
Here $\lambda_j$ are the eigenvalues of the reduced density operator of region $A_a$. When we
truncate after $D$ Schmidt coefficients in the cut corresponding to $A_a$, we will produce an error
in the description of the state
\begin{equation}
    t_a(D):=\sum_{j>D}\lambda_j.
    \label{eq:schmidtTail}
\end{equation}
The sequential procedure produces an unnormalized MPS $\ket{\phi_{D}}$, with every bond dimension
at most $D$, satisfying
\begin{equation}
    \bigl\|\ket{\omega}-\ket{\phi_{D}}\bigr\|_2^2 \leq \sum_{a=1}^{n-1}t_a(D).
    \label{eq:truncationError}
\end{equation}
Although the actual singular values encountered after the first truncation can differ from those in
Eq.~\eqref{eq:schmidtDecomposition}, the bound uses the tails of the original state. Indeed, the previously retained
tensors define an isometry from the retained left index into the physical left block. Contracting
with its adjoint cannot increase the best rank-$D$ approximation error at the next cut. Moreover,
the errors discarded at successive steps are orthogonal, so their squared norms add.

In Appendix~\ref{sec:appendixMPS} we prove the following lemma, which bounds the error produced at
each cut uniformly, i.e. independent of $a$.

\begin{lemma}\label{lem:chainTail}
    For $a=1,\ldots,n-1$,
    \begin{equation}
        \label{eq:chainTailBound}
        t_a(D)\leq\frac{C_0n^{2/\pi}}{D},\qquad C_0=e^{1+4/\pi}.
    \end{equation}
\end{lemma}

The proof uses explicitly that the ground state is a Gaussian state, so that the singular values
can be directly connected to properties of the matrix $G$ in Eq.~\eqref{eq:matrixG} when one considers
the same partition in the single-particle sector. Using Lemma~\ref{lem:chainTail} and the sequential procedure to
obtain the MPS described above one arrives at:

\begin{theorem}\label{thm:chainMPS}
    For every $\epsilon>0$, there is a normalized MPS $\ket{\omega_D}$ satisfying
    \begin{equation}
        \label{eq:mpsApproximation}
        \bigl\|\ket{\omega}-\ket{\omega_D}\bigr\|_2\leq\epsilon
    \end{equation}
    with bond dimension $D=$poly$(n,1/\epsilon)$.
\end{theorem}

\begin{proof}
    Lemma~\ref{lem:chainTail} and Eq.~\eqref{eq:truncationError} give
    \begin{equation}
    \label{eq:display007}
        \bigl\|\ket{\omega}-\ket{\phi_D}\bigr\|_2^2
        \leq\frac{C_0(n-1)n^{2/\pi}}{D}.
    \end{equation}
    Normalizing $\ket{\phi_D}$ increases the error by at most a factor of two. We take
    \begin{equation}
        \label{eq:mpsBondBound}
        D=\left\lceil\frac{4C_0(n-1)n^{2/\pi}}{\epsilon^2}\right\rceil
        =O\!\left(\frac{n^{1+2/\pi}}{\epsilon^2}\right),
    \end{equation}
    where $C_0=e^{1+4/\pi}$, which proves the claim.
\end{proof}

% ------------------------------------------------------
\subsection{MPU description of the wavelet operator}
\label{sec:waveletMPU}

As shown in Refs.~\cite{evenbly2016wavelets,haegeman2018wavelets}, one can find a unitary operator, $U$, that transforms $\boldsymbol\eta_n$ into $\boldsymbol\eta_{n/2}$ plus some other modes that are localized in the original lattice. This was used in order to obtain efficient descriptions of critical states in
terms of MPS by sequentially applying such transformations. Here we will be concerned with $U$
itself and show that it can be well approximated by an MPU with a bond dimension
$D=\operatorname{poly}(n,1/\epsilon)$. Whenever it is needed, we will use the subindex $n$ (or $n/2$) to denote
that the corresponding matrix acts on $n$ (or $n/2$) sites.

Let us define and characterize the wavelet unitary operator $U$. We define
$\boldsymbol c^e=(c_0,c_2,\ldots,c_{n-2})^T$ and $\boldsymbol c^o=(c_1,c_3,\ldots,c_{n-1})^T$ the vectors that
contain the $c$ operators in even and odd sites, respectively and similarly for $\boldsymbol d^e$
and $\boldsymbol d^o$. We can express
\begin{equation}
    \label{eq:display008}
    \boldsymbol c^e\oplus \boldsymbol c^o= Q \boldsymbol c,\qquad
    \boldsymbol d^e\oplus \boldsymbol d^o= Q \boldsymbol d
\end{equation}
where $\oplus$ indicates here that the vectors have to be concatenated, and $Q$ is a matrix, which
in ket notation can be written as
\begin{equation}
    \label{eq:evenOddPermutation}
    Q = \sum_{x\in \mathcal X_{n/2}} \ket{x}\bra{2x} + \ket{x+n/2}\bra{2x+1}
\end{equation}
In Appendix~\ref{sec:appendixMPU}, Sec.~\ref{sec:filterConstruction}, we introduce two $n\times n$ unitary matrices, $H$ and $J$
such that
\begin{equation}
    \label{eq:waveletIdentity}
    H^\dagger G_{n}J = G_{n/2}\oplus \Id_{n/2}.
\end{equation}
where we have retained the size of the matrix $G$ using the subindex, and $\Id_{n/2}$ is the identity
matrix acting on $n/2$ sites.

\begin{figure}[t]
    \centering
    \includegraphics[width=\columnwidth]{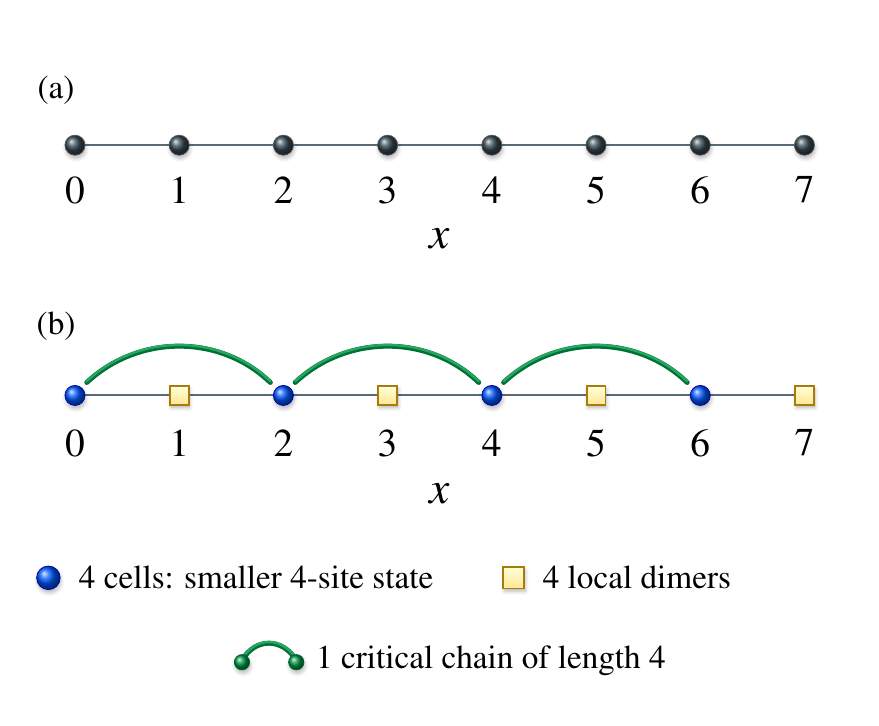}
    \caption{The first one-dimensional wavelet step for $n=8$, corresponding to Eq.~\eqref{eq:waveletOrbitals}. (a) The initial chain has eight cells, each containing the two modes $(c_x,d_x)$. (b) The four even-coordinate cells (blue circles) carry the smaller critical state $\ket{\omega_4}$, while the four odd-coordinate cells (gold squares) contain onsite dimers $(c_x^\dagger+d_x^\dagger)\ket0/\sqrt2$. The green curves connect successive cells of the same critical chain; they do not represent additional states or a sequence of two-mode gates. The periodic closing connection is omitted.}
    \label{fig:wavelet1D}
\end{figure}

Now, we can define the wavelet operator $U$ by its action on
\begin{subequations}
    \label{eq:waveletAction}
    \begin{align}
        U\boldsymbol c^\dagger U^\dagger&= U^c\boldsymbol c^\dagger U^{c\dagger}
        =\boldsymbol c^\dagger C_h^\dagger,\\
        U\boldsymbol d^\dagger U^\dagger
        &= U^d\boldsymbol d^\dagger U^{d\dagger}
        =\boldsymbol d^\dagger C_j^\dagger,
        \label{eq:waveletDAction}
        \end{align}
\end{subequations}
where
\begin{equation}
    \label{eq:waveletMatrices}
    C_h = HQ, \qquad C_j=JQ.
\end{equation}
Note that it is always possible to write
\begin{equation}
    \label{eq:waveletFactorization}
    U=U^c U^d
\end{equation}
where $U^c$ and $U^d$ act only on the $c$ and $d$ modes, respectively. Thus, we have
\begin{equation}
    U\boldsymbol\eta^\dagger U^\dagger J
    = \frac{\boldsymbol c^{e \dagger} G_{n/2}+\boldsymbol d^{e\dagger}}{\sqrt{2}}
    \;\oplus\;
    \frac{\boldsymbol c^{o\dagger}+\boldsymbol d^{o\dagger}}{\sqrt{2}}.
    \label{eq:waveletOrbitals}
\end{equation}
Thus, after applying $U$ we obtain a set of new modes that (up to linear combinations via the
unitary matrix $J^\dagger$), can be expressed as some local operators acting on the odd sites, and
the ones that generate the ground state of a chain of $n/2$ sites at the even positions. Fig.~\ref{fig:wavelet1D} illustrates the action of $U$ for $n=8$.

The matrices $H$ and $J$ are parametrized by $p\in\mathbb N$ and are block diagonal in momentum
representation. For $H$, the dependence of each block on $z=e^{ik}$ can be expressed as
quotients of polynomials of degree $2p$, a property which is inherited by $C_h$. As we show in
Appendix~\ref{sec:appendixMPU}, Sec.~\ref{sec:mpoBounds}, this immediately implies that $U^c$ can be written as an MPU with bond
dimension $D_c\le 2^{16p}$. The idea is to show that for any cut between a region $A$ and its
complement, the Schmidt rank of the corresponding Choi state is related to the rank of a submatrix
of $C_h$ associated to the same cut. Such rank can be easily bounded if the blocks of $H$ can
be written in terms of polynomials of $z$, since they correspond to translations in position
representation.

The matrix $J$ cannot be treated in the same way, since its momentum representation cannot be
written as a quotient of polynomials in $z$. In order to get around this obstacle, we approximate
$J$ by another matrix, $\tilde J$, which can be written in that form with polynomials of degree
$3p$. Thus, we replace $U^d$ by $\tilde U^d$, which is defined as in Eq.~\eqref{eq:waveletAction} but
with the replacement $J\to \tilde J$. Similar to $U^c$, this last operator has a bond dimension
$D_d\le 2^{24p}$. Defining $\tilde U=U^c\tilde U^d$, we obtain an exact MPU representation of bond
dimension $D\le 2^{40p}$, and an error (see Eq.~\eqref{eq:rationalUnitaryError})
\begin{equation}
    \label{eq:oneStepMPUError}
    \begin{aligned}
        \|\tilde U-U\|_{\mathrm{op}}=\|\tilde U^d-U^d\|_{\mathrm{op}}\leq2\sqrt2\,n\rho^p.
    \end{aligned}
\end{equation}
Thus, for a prescribed $0<\epsilon<1$, we can choose
\begin{equation}
    \label{eq:filterParameter}
    p=\left\lceil\frac{\log(2\sqrt2\,n/\epsilon)}{\log(1/\rho)}\right\rceil.
\end{equation}

We emphasize that both $U^c$ and $\tilde U^d$ are unitaries and Gaussian, and thus $\tilde U$
inherits both properties. This leads to the following result which is fully proved in Appendix~\ref{sec:appendixMPU}.
\begin{theorem}
    \label{thm:waveletMPU}
    The wavelet operator $U$ can be approximated up to error $\epsilon$ by a Gaussian unitary MPU with bond dimension $D=O((n/\epsilon)^{\kappa})$.
\end{theorem}
Using explicitly the bounds given in the lemmas of the Appendix, we can give a more precise bound
\begin{equation}
    \label{eq:waveletMPUBond}
    D\le 2^{40}\left(\frac{2\sqrt2\,n}{\epsilon}\right)^{\kappa},
    \qquad \kappa=\frac{40\log 2}{\log(1/\rho)},
    \qquad \rho=3-2\sqrt2.
\end{equation}

% ======================================================
\section{2D tight-binding model: PEPS approximation}
\label{sec:pepsDescription}

In this Section, we construct a PEPS approximation to the ground state of the two-dimensional
tight-binding models by using the one-dimensional MPUs constructed in Sec.~\ref{sec:oneDimensionalApproximations}.
At each step, we apply them along the rows and columns of the active lattice. This separates a
smaller copy of the two-dimensional state, horizontal and vertical critical chains, and local
dimers. We exchange the cells of the smaller system with dimers at prescribed positions, so that
different scales use different rows and columns. As we will show, this ensures that each row
and column is involved in at most two steps, and thus that the unitaries are only applied a
number of times independent of the system size. We then replace the critical chains by the MPSs of
Sec.~\ref{sec:groundStateMPS} and reverse the procedure, using the exact MPUs introduced in
Sec.~\ref{sec:rationalApproximation}. The bond dimension is controlled by counting the
MPUs, MPS, and SWAP factors on each nearest-neighbor edge. We will first treat the two-mode per
cell model $\tilde H_2$ and finally obtain a PEPS for the specified ground state of $H_2$. We
include subindices $n$ to denote the size of the lattice we are considering at each step so, for
instance, $\ket{\Omega_n}$ is the ground state in a lattice with $n\times n$ sites.

% ------------------------------------------------------
\subsection{Construction}
\label{sec:pepsConstruction}

We consider an $n\times n$ periodic lattice for $\tilde H_2$ in Eq.~\eqref{eq:auxiliaryHamiltonian2D},  with
$n=2m=2^\ell$. Let $U_{y}^{c,\mathrm{row}}$ be the same unitary as $U^c$ of Sec.~\ref{sec:waveletMPU}, now acting on $(c_{0,y},\ldots,c_{n-1,y})$, and let
$U_{x}^{c,\mathrm{col}}$ act on $(c_{x,0},\ldots,c_{x,n-1})$. We define the analogous operators for
$d$. The wavelet unitary operator
\begin{subequations}
    \label{eq:rowColumnUnitary}
    \begin{align}
    \label{eq:display009}
        W&=W^{c}W^{d},\\
        W^{\sigma} &=\left(\prod_{x\in\mathcal X}U_{x}^{\sigma,\mathrm{col}}\right)
          \left(\prod_{y\in\mathcal X}U_{y}^{\sigma,\mathrm{row}}\right),
          \quad\sigma=c,d.
    \end{align}
\end{subequations}

We arrange the operators $c_{x,y}$ and $d_{x,y}$ as in Sec.~\ref{sec:freeFermionModels}. The action of $W$ on
the modes is precisely the tensor product of the one-dimensional actions:
\begin{subequations}
    \label{eq:rowColumnAction}
    \begin{align}
    \label{eq:display010}
        W\boldsymbol c^\dagger W^{\dagger}
        &=\boldsymbol c^\dagger(C_h^\dagger\otimes C_h^\dagger),\\
        W\boldsymbol d^\dagger W^\dagger
        &=\boldsymbol d^\dagger(C_j^\dagger\otimes C_j^\dagger).
    \end{align}
\end{subequations}
For example, the first equation means
\begin{equation}
    \label{eq:display011}
    Wc_{x,y}^\dagger W^{\dagger}
    =\sum_{a,b\in\mathcal X}(C_h^\dagger)_{a,x}(C_h^\dagger)_{b,y}c_{a,b}^\dagger.
\end{equation}
Note that $W^c$ and $W^d$ are Gaussian transformations and they commute with each other since they
only involve $c$ and $d$ operators. Furthermore, the product of all unitaries for $c$ (or $d$)
acting on rows commutes with the product of the unitaries acting on the columns since their
one-particle matrices act on different coordinates. However, an individual row unitary need not
commute with an individual column unitary of the same species.

As in Eq.~\eqref{eq:waveletOrbitals}, we can easily describe the action of the wavelet unitary on
$\boldsymbol \eta$. For that, it is convenient to define
\begin{equation}
    \label{eq:orbitalAction2D}
    \boldsymbol\zeta^\dagger
    =W\boldsymbol\eta^\dagger W^{\dagger}(J\otimes J).
\end{equation}

For $\mu,\nu\in\{0,1\}$, let $\boldsymbol c^{\mu\nu}$ contain the modes $c_{2r+\mu,2s+\nu}$,  and
similarly for $d$. Let $\boldsymbol\zeta^{\mu\nu}$ contain the components of $\boldsymbol\zeta$
with labels $(r+\mu m,s+\nu m)$, in the same order in which we sorted the rest of the operators to
form vectors in 2D, i.e., first rows and then columns. Then Eq.~\eqref{eq:waveletIdentity} yields
\begin{equation}
    \label{eq:orbitalSectors2D}
    \boldsymbol\zeta^{\mu\nu\dagger}
    =\frac{\boldsymbol c^{\mu\nu\dagger}(A_\mu\otimes A_\nu)
     +\boldsymbol d^{\mu\nu\dagger}}{\sqrt2},
\end{equation}
where $A_0=G_m$, and $A_1=\Id_m$. Thus, after listing the four groups in the order $00,01,10,11$,
the matrix defining the occupied modes is the direct sum
\begin{equation}
    \label{eq:directSums2D}
    \begin{aligned}
        &(G_m\otimes G_m)\oplus(G_m\otimes\Id_m)\\
        &\hspace{10mm}\oplus(\Id_m\otimes G_m)\oplus\Id_{m^2}.
    \end{aligned}
\end{equation}
The multiplication by $J\otimes J$ in Eq.~\eqref{eq:orbitalAction2D} acts only on the occupied-orbital
labels and as it was mentioned above, does not carry any implication.

\begin{figure}[t]
    \centering
    \includegraphics[width=\columnwidth]{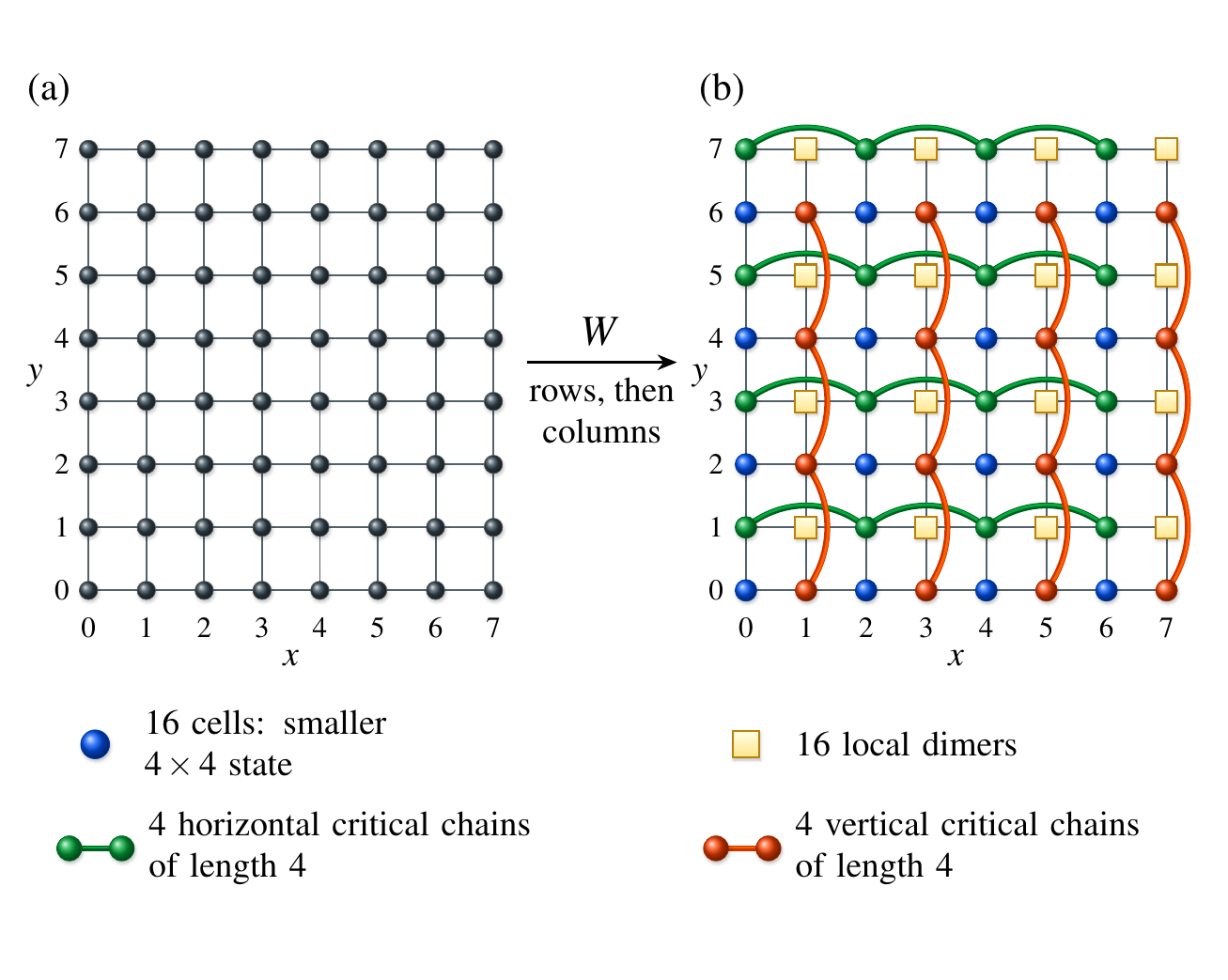}
    \caption{The first wavelet step on an $8\times8$ lattice of two-mode cells. (a) The initial lattice. (b) The resulting four sectors: 16 even--even cells (blue circles) carry $\ket{\Omega_4}$; green curves indicate four horizontal critical chains; orange curves indicate four vertical critical chains; and 16 gold squares denote local dimers. The symbol $W$ in the illustration denotes the row-and-column unitary of Eq.~\eqref{eq:rowColumnUnitary}. Curves show the placement of independent state factors, not finite-range gates. No SWAP operation is needed at this first step.}
    \label{fig:split8}
\end{figure}

If we apply $W$ to the ground state, the lattice can be divided into four groups of sites
corresponding to $\mu,\nu=0,1$:
\begin{equation}
    \label{eq:sectorPositions2D}
    \scalebox{0.975}{$\displaystyle
    \begin{array}{@{}l@{\,}l@{}}
        (2r,2s)&\text{two-dimensional state in an $m\times m$ lattice},\\
        (2r,2s+1)&\text{horizontal chain, with }s\text{ fixed},\\
        (2r+1,2s)&\text{vertical chain, with }r\text{ fixed},\\
        (2r+1,2s+1)&\text{local dimer},
    \end{array}$}
\end{equation}
with $r,s\in\mathcal X_m$.  The first group is $\ket{\Omega_m}$. The second consists of $m$
independent copies of the length-$m$ state $\ket{\omega_m}$ of Sec.~\ref{sec:oneDimensionalApproximations}, and
the third gives another $m$ copies. The last group is a product of the states
\begin{equation}
    \label{eq:localDimer}
    \ket{D_{x,y}}=\frac{c_{x,y}^\dagger+d_{x,y}^\dagger}{\sqrt2}\ket0.
\end{equation}
Consequently, up to a global phase and with these positions understood,
\begin{equation}
    \label{eq:stateSplitting2D}
    W\ket{\Omega_n}
    =\ket{\Omega_m}\otimes\ket{\omega_m}_{\mathrm h}^{\otimes m}
       \otimes\ket{\omega_m}_{\mathrm v}^{\otimes m}\otimes\ket D^{\otimes m^2}.
\end{equation}
Note that every tensor factor has a definite particle number (at half filling) and occupies a
disjoint set of modes. In Fig.~\ref{fig:split8} we have illustrated the action of $W$ in an
$8\times 8$ lattice.

In principle, we could continue now and apply $W$ to the smaller system containing a copy of the original state but in an $n/2\times n/2$ lattice. However, this would give rise to a
superpolynomial bond dimension, since we would be applying the operation $W$ to some of the same rows and columns at each step. In order to avoid that, we move the active lattice so that in each step the operation $W$ acts on different rows and columns. For that, we use a series of fermionic swap operations, $S$, that take each active site in the lattice to one where there is a dimer. Those swap operations increase the bond dimension; however, again they are chosen so that they never affect the same bonds at each step, so that this will provide a constant multiplicative factor. Note that they are Gaussian as well, so that the whole procedure keeps its Gaussian nature.

At step zero, Eq.~\eqref{eq:sectorPositions2D} already places the smaller state on the even--even cells. At the next step, the Gaussian transformations act on the even-coordinate lattice. They leave the next smaller state at $(4r,4s)$, whereas its desired positions are $(4r+1,4s+1)$. The latter cells already contain dimers from step zero. We thus swap each of these pairs of cells so that the dimers are moved to the positions $(4r,4s)$ and the smaller state is now in the desired positions.

We can specify the position of the active cells at each step as follows. Let $n_j=n/2^j$ and
\begin{subequations}
    \label{eq:activeCoordinates}
    \begin{align}
    \label{eq:display012}
        x_0(t)&=t,\\
        x_j(t)&=2^jt+2^{j-1}-1,\qquad j\geq1,\\
        \mathcal X^{(j)}&=\{x_j(t):t\in\mathcal X_{n_j}\}.
    \end{align}
\end{subequations}
The active system at step $j$ will be on the lattice defined by $\mathcal X^{(j)}\times\mathcal
X^{(j)}$. It is easy to check that the sets $\mathcal X^{(j)}$ for $j\geq1$ are disjoint: if
$x\in\mathcal X^{(j)}$, then $x+1$ has exactly $j-1$ factors of two. Thus, as announced, at each
step (except for the first one), the active lattice uses different rows and columns. In Appendix~\ref{sec:appendixPEPS}, Sec.~\ref{sec:fermionicSWAPs}, we explain in detail the procedure.

\begin{figure*}[t]
    \centering
    \includegraphics[width=\textwidth]{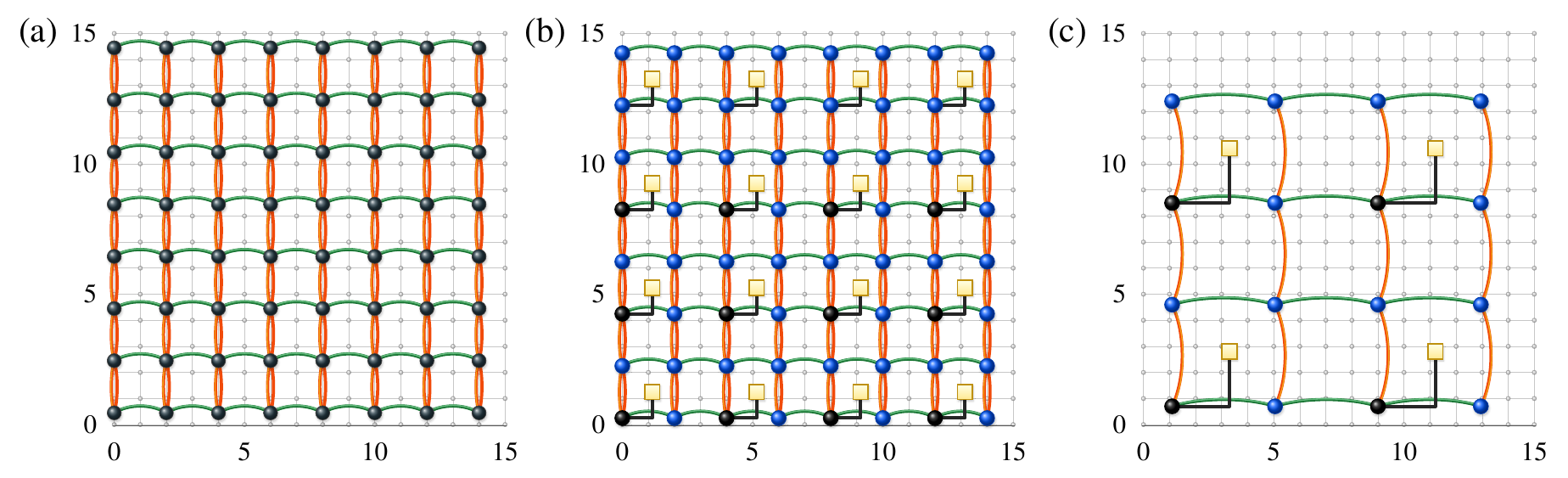}
    \caption{Three steps of the multiscale construction on a $16\times16$ lattice. (a) Step $W_0$: horizontal and vertical curvy lines indicate the row and column wavelet actions on the full lattice, while the black circles mark the active lattice $\mathcal X^{(1)}\times \mathcal X^{(1)}$. (b) Step $W_1$: the blue circles show the incoming active lattice $\mathcal X^{(1)}\times \mathcal X^{(1)}$; the black circles mark the smaller lattice produced by $W_1$, namely $\{0,4,8,12\}^2$; the thick L-shaped segments show the swap paths of Eq.~\eqref{eq:swapPath}; and the gold squares mark the new active lattice $\mathcal X^{(2)}\times \mathcal X^{(2)}$. (c) Step $W_2$: the same conventions are used for the incoming active lattice $\mathcal X^{(2)}\times \mathcal X^{(2)}$, the intermediate lattice $\{1,9\}^2$, and the final active lattice $\mathcal X^{(3)}\times \mathcal X^{(3)}$. The exact active-coordinate sets are defined in Eq.~\eqref{eq:activeCoordinates}, and the swap endpoints are given in Eq.~\eqref{eq:swapEndpoints}.}
    \label{fig:procedure16}
\end{figure*}

For example, at $n=16$ the active coordinates are
\begin{equation}
    \label{eq:activeCoordinates16}
    \begin{aligned}
        \mathcal X^{(1)}&=\{0,2,\ldots,14\},\\
        \mathcal X^{(2)}&=\{1,5,9,13\},\qquad
        \mathcal X^{(3)}=\{3,11\}.
    \end{aligned}
\end{equation}
The first 16 exchanges are $(4r,4s)\leftrightarrow(4r+1,4s+1)$,  with $0\leq r,s<4$. The next four
are $(8r+1,8s+1)\leftrightarrow(8r+3,8s+3)$, with $0\leq r,s<2$. Fig.~\ref{fig:procedure16}
displays both the wavelet steps and these exchanges. The recursion stops at side two, because
$G_2=\Id_2$ and $\ket{\Omega_2}$ is already a product of four dimers.

The complete exact analysis is, with the rightmost factor acting first,
\begin{equation}
    \label{eq:completeAnalysis2D}
    \mathcal V=P_{\ell-2}W_{\ell-2}\cdots P_1W_1W_0.
\end{equation}
Its action gives the resource state
\begin{equation}
    \label{eq:resource2D}
        \mathcal V\ket{\Omega_n} =\ket{\Omega_2}\otimes
        \bigotimes_{m=n/2,n/4,\ldots,2}
        \left(\ket{\omega_m}^{\otimes2m}\otimes\ket D^{\otimes m^2}\right).
\end{equation}
The positions of the factors  are the ones just specified.

A row or column carries at most one 1D wavelet operator, since the sets $\mathcal X^{(j)}$, $j\geq1$, are disjoint. Including the first step, $W_0$, gives at most two 1D wavelet operators. Note that each edge in the original lattice is used by at most two SWAP gates.

% ------------------------------------------------------
\subsection{Bond-dimension and error}
\label{sec:pepsBounds}

The PEPS is constructed starting out with a state composed of the original dimers and the critical
chains in Eq.~\eqref{eq:resource2D}, but where those are replaced by the MPS approximations of Sec.~\ref{sec:oneDimensionalApproximations}. Then, we construct the PEPS by applying $\mathcal V^\dagger$
from Eq.~\eqref{eq:completeAnalysis2D} but where we replace every wavelet transformation acting on the $d$
modes by $\tilde U^d$. In this subsection we will compute the bond dimension and the error, and fix
the parameter $p$ in order to obtain the desired scaling.

Let us call $D_1$ the bond dimension of the normalized MPS in Theorem~\ref{thm:chainMPS} to
distinguish it from the final PEPS bond dimension $D$.

We now approximate each critical chain in Eq.~\eqref{eq:resource2D} by the normalized MPS of
Theorem~\ref{thm:chainMPS} and keep all dimers intact. Let us denote by
$\ket{\tilde\Phi_{D_1}}$ that state, where $D_1$ is the bond dimension, which is different from the
final PEPS bond dimension $D$. By Theorem~\ref{thm:chainMPS}, using $(m-1)m^{2/\pi}\leq m^2$, replacing
$\ket{\omega_m}$ by the corresponding MPS gives
\begin{equation}
    \label{eq:chainErrorPEPS}
    \bigl\|\ket{\omega_m}-\ket{\omega_{m,D_1}}\bigr\|_2^2
       \leq\frac{4C_0m^2}{D_1},\qquad C_0=e^{1+4/\pi}.
\end{equation}
Note that the Schmidt vectors may be chosen in fixed particle-number sectors, so the truncated MPS
has the same particle number and definite fermionic parity.

In each $W_j$ in Eq.~\eqref{eq:completeAnalysis2D}, we replace each $d$-mode 1D wavelet transformation
by $\tilde U^d$ of Lemma~\ref{lem:dModeMPO}, retaining the $c$-mode transformations exactly. We
denote this step by $\tilde W_j$ and use the same SWAP $P_j$ to define
\begin{subequations}
    \label{eq:approximatePEPS}
    \begin{align}
    \label{eq:display013}
        \tilde{\mathcal V}_{p}
        &=P_{\ell-2}\tilde W_{\ell-2}\cdots P_1\tilde W_1\tilde W_0,\\
        \ket{\tilde\Omega_n}
        &=\tilde{\mathcal V}_{p}^{\dagger}\ket{\tilde\Phi_{D_1}}.
    \end{align}
\end{subequations}
In this inverse construction we undo a cell permutation and then apply the inverse Gaussian row and
column transformations at that scale, proceeding from coarse to fine. Every applied operator is
exactly unitary, and the only non-Gaussian part corresponds to the MPS resource. The resulting
state is therefore normalized. We thus have:

\begin{lemma}
    \label{lem:pepsBond}
    The state $\ket{\tilde\Omega_n}$ has a fermionic PEPS representation on the two-mode square lattice
    with bond dimension
    \begin{equation}
        \label{eq:coverBond}
        D \leq 256D_1^2\,2^{80p}.
    \end{equation}
\end{lemma}
The proof is given in Appendix~\ref{sec:appendixPEPS}, Sec.~\ref{sec:appendixPEPSProofs}, and uses a simple counting based on the
construction above.

In the same appendix, we compute the error
\begin{lemma}
    \label{lem:pepsError}
    The PEPS $\ket{\tilde\Omega_n}$ obeys
    \begin{equation}
        \label{eq:coverError}
        \bigl\|\ket{\tilde\Omega_n}-\ket{\Omega_n}\bigr\|_2
        \leq8n^2\rho^p+
        \sqrt{\frac{8C_0(n^3-8)}{7D_1}},
    \end{equation}
    where $\rho=3-2\sqrt2$.
\end{lemma}

With those lemmas, we can easily arrive at the final result

\begin{theorem}
    \label{thm:auxiliaryPEPS}
    Let $n=2^\ell\geq4$ and $0<\epsilon<1$. There exists a normalized fermionic PEPS
    $\ket{\tilde\Omega_n}$ such that
    \begin{equation}
        \label{eq:auxiliaryPEPSBound}
        \bigl\|\ket{\tilde\Omega_n}-\ket{\Omega_n}\bigr\|_2
        \leq\epsilon,
    \end{equation}
    with bond dimension $D=\operatorname{poly}(n,1/\epsilon)$.
\end{theorem}

\begin{proof}
We choose $p$ and $D_1$ as
   \begin{equation}
        \label{eq:auxiliaryPEPSParameters}
        p=\left\lceil \frac{\log(16n^2/\epsilon)}
        {\log(1/\rho)}\right\rceil,\qquad
        D_1=\left\lceil         \frac{32C_0(n^3-8)}{7\epsilon^2}
        \right\rceil,
     \end{equation}
with $\rho=3-2\sqrt2$ and $C_0=e^{1+4/\pi}$. Each term in the error bound of Lemma~\ref{lem:pepsError} is then at most $\epsilon/2$, while Lemma~\ref{lem:pepsBond}
gives $D=O\left(n^{6+2\kappa}/\epsilon^{4+\kappa}\right)$, and
    $\kappa=80\log2/\log(1/\rho)$.
\end{proof}

% ------------------------------------------------------
\subsection{PEPS approximation for the tight-binding Hamiltonian}
\label{sec:physicalPEPS}

The preceding bounds concern the two-mode model $\tilde H_2$. To state the result also for the one-mode Hamiltonian $H_2$, we use the rotated coordinates introduced in Ref. \cite{haegeman2018wavelets}. In Appendix~\ref{sec:appendixH2} we show that using those coordinates, folding the PEPS, extending the physical dimension and re-routing the bonds, the PEPS approximation to $\ket{\Omega_n}$ allows us to directly build another PEPS approximation for the tight-binding model.

\begin{lemma}
    \label{lem:physicalExtraction}
    There is a specified half-filled periodic ground state $\ket{\Omega_n^{\mathrm{per}}}$ of $H_2$ with the following property. A normalized PEPS approximation to $\ket{\Omega_n}$ of bond $D_{\mathrm{cov}}$ and error $\delta<1$ gives a normalized one-mode PEPS of error at most $2\delta$ and bond at most $4D_{\mathrm{cov}}^4$.
\end{lemma}

We can now obtain the PEPS approximation for the original tight-binding model, which immediately follows from Theorem~\ref{thm:auxiliaryPEPS} and Lemma~\ref{lem:physicalExtraction}.
\begin{theorem}
    \label{thm:pepsDescription}
    Let $n=2^\ell\geq4$ and $0<\epsilon<1$. There exists a normalized
    one-mode fermionic PEPS $\ket{\tilde\Omega_n^{\mathrm{per}}}$ satisfying
    \begin{equation}
        \label{eq:physicalPEPSBound}
        \bigl\|\ket{\tilde\Omega_n^{\mathrm{per}}}
              -\ket{\Omega_n^{\mathrm{per}}}\bigr\|_2\leq\epsilon
    \end{equation}
    for the specified periodic ground state of $H_2$, with bond dimension
    $D=\operatorname{poly}(n,1/\epsilon)$.
\end{theorem}

Note that all the bounds are conservative and are intended to establish polynomial scaling. Note also that the critical-chain MPS approximations, and therefore the final PEPS, need not be Gaussian.

% ======================================================
% ======================================================
\section{2D Gapped Gaussian models}
\label{sec:gappedGaussianPEPS}

In this Section, we consider ground states of gapped chiral Gaussian models, such as the Chern insulator introduced in Sec.~\ref{sec:gappedGaussianModels} and $p+ip$ superconductors. These states are ground states of local quadratic Hamiltonians of the form in Eq.~\eqref{eq:generalGaussianHamiltonian}. We will show that all such uniformly gapped, finite-range models admit Gaussian PEPS approximations with bond dimension polynomial in the system size and the inverse error.

For the Chern insulator in Eq.~\eqref{eq:chernHamiltonian}, the construction can be made particularly explicit by approximating the square root in its occupied orbitals. We retain that construction in Appendix~\ref{sec:appendixChern}. The argument given here applies more generally and does not require translation invariance or an explicit expression for the occupied orbitals.

We first consider particle-number-conserving models. The idea is to start with a suitably chosen occupation product state and apply a fixed-range one-particle filter repeatedly. The normalized state approaches the ground state exponentially with the number of repetitions. We then extend the result to general Gaussian Hamiltonians by taking two copies and making a local change of fermionic variables.

Although we are particularly interested in two dimensions, the argument applies in any fixed spatial dimension. The constants depend on the dimension, the interaction range, the gap, and the fixed number of modes per site, but not on the lattice size or the requested error. In two dimensions $N=O(n^2)$, so polynomial bounds in $N$ are also polynomial in $n$.

% ------------------------------------------------------
\subsection{Particle number conserving models}
\label{sec:gappedNumberConserving}

We consider the Hamiltonian in Sec.~\ref{sec:gappedGaussianModels} with $G=0$, satisfying Eq.~\eqref{eq:gappedSpectrum}: This condition together with Eq.~\eqref{eq:h_in_terms_of_JG} as well as the normalization assumption in Eq.~\eqref{eq:normalization} yields that
\begin{align}\label{eq:spec_condition_J}
    \Delta^2  \Id\preceq J^\dagger J \prec \Id.
\end{align}
In this case, the ground state is obtained by filling all negative-energy orbitals:
\begin{equation}
    \label{eq:gappedGroundState}
    \ket{\Phi}=\prod_{i=1}^{r}{\boldsymbol{a}}^\dagger v_i\ket0,
\end{equation}
where $\boldsymbol{a}^\dagger = (a_1^\dagger, \dots, a_N^\dagger)$, $r$ is the number of negative eigenvalues of $J$ and $v_1,\ldots,v_r$ are orthonormal eigenvectors of $J$ with negative eigenvalues. Let $w_1,\ldots,w_{N-r}$ be orthonormal eigenvectors with positive eigenvalues, and denote by $V$ and $W$ the matrices containing these vectors as columns: By construction, $V$ and $W$ satisfy $V^\dagger V=\Id$, $W^\dagger W=\Id$, and $V^\dagger W=0$.

Note that in the cases $r=0$ and $r=N$, $\ket{\Phi}$ is already a product state, so in the following analysis we take $0<r<N$. To prepare $\ket{\Phi}$, we start with an occupation product state
\begin{equation}
    \label{eq:gappedReferenceState}
    \ket{\Phi_S}=\prod_{i\in S}a_i^\dagger\ket0,
    \qquad S\subseteq\{1,\ldots,N\},\quad |S|=r.
\end{equation}
All products use the same fixed ordering. Let $E_S$ be the $N\times r$ matrix whose columns are the coordinate vectors $e_i$, $i\in S$. We will choose $S$ such that $V^\dagger E_S$ is invertible and $\norm{(V^\dagger E_S)^{-1}}_\text{op}\leq \sqrt{r(N - r + 1)}$ i.e., the minimum singular value of $V^\dagger E_S$ is only polynomially small in $N, r$. In Appendix~\ref{sec:appendixGapped}, Sec.~\ref{sec:gappedFilterProofs}, we show that there always exist a choice of $S$ which satisfies this condition.

Let us define the number-conserving quadratic operator
\begin{equation}
    \label{eq:gappedFilterGenerator}
    K=\sum_{i,j}[\log(1-J)]_{ij}a_i^\dagger a_j,
\end{equation}
where the matrix logarithm is well defined because, from Eq.~(\ref{eq:spec_condition_J}), $\Id-J$ is strictly positive. For an integer $q\geq0$, set
\begin{equation}
    \label{eq:gappedFilteredState}
    \ket{\Phi_q}=\frac{e^{qK}\ket{\Phi_S}}
    {\|e^{qK}\ket{\Phi_S}\|}.
\end{equation}
Note that $e^K$ amplifies negative-energy modes relative to positive-energy modes. Indeed, it is easy to see that for any $u$,
\begin{equation}
    \label{eq:gappedFilterAction}
    e^K (\boldsymbol{a}^\dagger u)e^{-K}=\boldsymbol{a}^\dagger (\Id-J)u,
    \qquad e^K\ket0=\ket0,
\end{equation}
and therefore
\begin{equation}
    \label{eq:gappedFilteredOrbitals}
    e^{qK}\ket{\Phi_S}
    =\prod_{i \in S}\boldsymbol{a}^\dagger(\Id-J)^q e_{i}\ket0.
\end{equation}
Our main result in this subsection shows that a Gaussian PEPs with polynomial bond-dimension can be obtained by approximating $\ket{\Phi}$ with $\ket{\Phi_q}$ for a sufficiently large $q$.
\begin{theorem}
    \label{thm:gappedNumberConserving}
    Let $H$ be a finite-range particle-number-conserving Gaussian Hamiltonian satisfying Eq.~\eqref{eq:gappedSpectrum}, in any fixed spatial dimension. For every $0<\epsilon<1$,
    \begin{equation}
        \label{eq:gappedNumberConservingApproximation}
        \|\ket{\Phi_q}-e^{i\alpha}\ket{\Phi}\|\leq\epsilon\quad \text{for}\quad q \leq O(\log N) + O(\log \epsilon^{-1}).
    \end{equation}
    for some $\alpha \in [0, 2\pi)$. Furthermore, $\ket{\Phi_q}$ can be represented as a Gaussian PEPs with bond-dimension $\exp(O(q))$ and thus furnishes a Gaussian PEPs with bond-dimension $O(\textnormal{poly}(N, 1/\epsilon))$ approximating $\ket{\Phi}$ to error $\epsilon$. 
\end{theorem}
\begin{proof}
In Appendix~\ref{sec:appendixGapped} we show that, for a suitable $S$, the normalized action of $e^{qK}$ becomes an increasingly good ground-state filter and satisfies
\begin{equation}
    \label{eq:gappedFilterError}
    \|\ket{\Phi_q}-e^{i\alpha}\ket{\Phi}\|
    \leq r\sqrt{N-r+1}\,\bigg(\frac{1-\Delta}{1+\Delta}\bigg)^q,
\end{equation}
for some $\alpha$. Furthermore, while $K$ is not necessarily a geometrically local operator, the action of $e^{K}$ on the fermionic modes (Eq.~\eqref{eq:gappedFilteredOrbitals}) is geometrically local. Using this, we can show that $e^K$ has an exact Gaussian PEPO representation with a bond dimension $D_0$ dependent only on the interaction range of $H$ (Lemma~\ref{lem:gappedStepPEPO}). Applying $q$ such PEPOs to the product state in Eq.~\eqref{eq:gappedReferenceState} gives $D\leq D_0^q$. Thus, taking
\begin{equation}
    \label{eq:gappedFilterDegree}
    q=\left\lceil
    \frac{\log(r\sqrt{N-r+1}/\epsilon)}
    {\log((1+\Delta)/(1-\Delta))}\right\rceil,
\end{equation}
ensures that $\|\ket{\Phi_q}-e^{i\alpha}\ket{\Phi}\| \leq \epsilon$ and yields a bond-dimension $D \leq \text{poly}(N, 1/\epsilon)$.
\end{proof}

% ------------------------------------------------------
\subsection{General Gaussian gapped models}
\label{sec:gappedGeneralModels}

We next consider the case of $G \neq 0$ i.e., we allow pairing terms in Eq.~\eqref{eq:generalGaussianHamiltonian}. It will be convenient to work in the Majorana representation (Eq.~\eqref{eq:gaussian_H_majorana}): Taking two copies of the system and consider the Hamiltonian
\begin{align}
    \tilde{H} &= {H}\otimes \Id + \Id \otimes {H} - \text{Tr}(J)\Id \nonumber \\
    &=\sum_{i, j}\sum_{\alpha, \beta}h^{\alpha, \beta}_{i, j}(\gamma_i^\alpha \gamma_j^\beta + \eta_i^\alpha \eta_j^\beta),
\end{align}
where we denote by $\eta_i^\alpha$ the Majorana operators for the second copy. Define the operator 
\begin{align}\label{eq:change_variables}
    c_i^\alpha = \frac{1}{\sqrt{2}}(\gamma_i^\alpha + i \eta_i^\alpha),
\end{align}
it can be checked that $\{c_i^\alpha, c_j^\beta\} = 0$ and $\{(c_i^\alpha)^\dagger, c_j^\beta\} = \Id \delta_{i, j}\delta_{\alpha, \beta}$. Thus, $c_i^\alpha$ is an annihilation operator on the doubled system. Furthermore, in terms of $c_i^\alpha$, we have that
\begin{align}
    \tilde{H} = 2\sum_{i, j}\sum_{\alpha, \beta}  h_{i, j}^{\alpha, \beta}(c_i^\alpha)^\dagger c_j^\beta,
\end{align}
which is a finite-range geometrically local model satisfying the gap condition of the previous subsection. The ground state of $\tilde{H}$ can thus be approximated as a PEPs using Theorem~\ref{thm:gappedNumberConserving}.

Undoing the on-site rotation corresponding to the change of variables in Eq.~\eqref{eq:change_variables} gives a Gaussian PEPS close to $\ket{\Omega}\ket{\Omega}$, where $\ket{\Omega}$ is the original ground state. In Appendix~\ref{sec:appendixGapped}, Sec.~\ref{sec:gappedProjectionProof}, we show that one can project the second copy onto an occupation product state without increasing the bond dimension, and with at most a factor of two in the normalized error. The selected projection may have a small probability and thus no inverse probability enters the error estimate.

\begin{theorem}
    \label{thm:gappedGeneralGaussian}
    Let $H$ be a finite-range Gaussian Hamiltonian of the form in Eq.~\eqref{eq:generalGaussianHamiltonian}, with bounded coefficients and a uniform single-particle gap, in a fixed spatial dimension. For every $0<\epsilon<1$, its ground state $\ket{\Omega}$ has a normalized fermionic Gaussian PEPS approximation satisfying
    \begin{equation}
        \label{eq:gappedGeneralApproximation}
        \|\ket{\widetilde\Omega}-\ket{\Omega}\|\leq\epsilon,
        \qquad D=\operatorname{poly}(N,1/\epsilon).
    \end{equation}
\end{theorem}
\begin{proof}
We apply Theorem~\ref{thm:gappedNumberConserving} to the doubled Hamiltonian with error $\epsilon/2$, undo the onsite change of variables, and use Lemma~\ref{lem:gappedOneCopyProjection}. All additional operations are onsite, so the polynomial bond bound is unchanged.
\end{proof}

% ======================================================
\section{2D Gutzwiller projected models}
\label{sec:GutzwillerProjected}

As discussed in Sec.~\ref{sec:gutzwillerModels}, many chiral topological states can be constructed by projecting several copies of chiral Gaussian states. If each copy has an exact PEPS representation with bond dimension $D$, a local projection of $m$ copies gives a PEPS with bond dimension at most $D^m$. In our case, however, the Gaussian states are only approximated by PEPS. The norm of the projected vector can be exponentially small, so a generic error estimate before normalization could be exponentially amplified and ruin the polynomial approximation.

In this Section we show that this amplification can be avoided for the family in Eq.~\eqref{eq:gutzPowerState}, with a fixed number of copies. The essential point is that nearby Slater determinants have nearby occupied subspaces. We can pair the occupied and unoccupied modes so that their difference is described by small two-mode rotations. Controlling those rotations in the physical occupation basis gives only a polynomial amplification of the error. In particular, we will not need a lower bound on the norm of the projected vector.

We allow arbitrary filling: let $\ket{\Phi}$ have $r$ particles in $N$ modes. Here $m$ is the number of copies, not the particle number. The local map is the equal-occupancy map in Eq.~\eqref{eq:gutzLocalMap}; for $m=2$ it includes the particle--hole-transformed Chern-band construction~\cite{zhang2011topological,zhang2013nonabelian}. Our result concerns these power-state projections, rather than every possible local constraint on several Gaussian states.

\begin{theorem}
    \label{thm:gutzwillerPEPS}
    Let $\ket{\Phi}$ be a normalized particle-number-conserving Gaussian state with $r$ particles in $N$ modes, and let $m\geq2$ be fixed. Suppose that, for every $0<\epsilon<1$, it admits a normalized $r$-particle Gaussian PEPS approximation $\ket{\tilde\Phi}$ satisfying
    \begin{equation}
        \label{eq:gutzInputApprox}
        \|\ket{\tilde\Phi}-\ket{\Phi}\|\leq\epsilon,
        \qquad D=\operatorname{poly}(N,1/\epsilon).
    \end{equation}
    Then the state $\ket{\Psi}$ defined in Eq.~\eqref{eq:Psi0} admits, for every $0<\tilde\epsilon<1$, a normalized PEPS approximation $\ket{\tilde\Psi}$ satisfying
    \begin{equation}
        \label{eq:gutzOutputApprox}
        \|\ket{\tilde\Psi}-\ket{\Psi}\|\leq\tilde\epsilon,
        \qquad \tilde D=\operatorname{poly}(N,1/\tilde\epsilon).
    \end{equation}
\end{theorem}
\begin{proof}
Let $\ket{\tilde\Phi}$ be a Gaussian PEPS satisfying Eq.~\eqref{eq:gutzInputApprox}. Analogously to Eq.~\eqref{eq:Psi0}, we define
\begin{equation}
    \label{eq:tildePsi}
    \ket{\tilde\Psi_0}=\left(\prod_{i=1}^{N}P_i\right)\ket{\tilde\Phi}^{\otimes m},
    \qquad
    \ket{\tilde\Psi}=\frac{\ket{\tilde\Psi_0}}{\|\ket{\tilde\Psi_0}\|}.
\end{equation}
The projected vectors are nonzero, since they contain the $m$th powers of the nonzero occupation amplitudes.

Since $\ket{\Phi}$ and $\ket{\tilde\Phi}$ are particle-number-conserving pure Gaussian states with $r$ particles, they are Slater determinants. Hence
\begin{equation}
    \label{eq:PhitildePhi}
    \ket{\Phi}=\prod_{i=1}^{r}\alpha_i^\dagger\ket0,
    \qquad
    \ket{\tilde\Phi}=\prod_{i=1}^{r}\tilde\alpha_i^\dagger\ket0.
\end{equation}
We now define the occupied-orbital overlap matrix as
\begin{equation}
    \label{eq:gutzOverlapMatrix}
    B_{ij}=\{\alpha_i,\tilde\alpha_j^\dagger\},
    \qquad B=W\Sigma V^\dagger,
\end{equation}
where $\Sigma$ contains the singular values $s_i$. Rotations within the occupied subspaces change the Slater determinants only by overall phases, which we fix consistently. After reordering, let $t$ be the number of singular values smaller than one. The residual components of the corresponding modes in the empty subspace are orthonormal, so $t\leq\min\{r,N-r\}$. We can therefore choose canonical modes $\beta_1,\ldots,\beta_N$ such that
\begin{equation}
    \label{eq:tildebetai}
    \begin{aligned}
        \tilde\beta_i&=s_i\beta_i+\sqrt{1-s_i^2}\,\beta_{r+i},
        &&1\leq i\leq t,\\
        \tilde\beta_i&=\beta_i,&&t<i\leq r,
    \end{aligned}
\end{equation}
with $\beta_{r+i}\ket{\Phi}=0$. Thus only the modes that actually change need an empty partner, including when the filling is larger than one half. In this basis,
\begin{equation}
    \label{eq:PhitildePhibeta}
    \ket{\Phi}=\prod_{i=1}^{r}\beta_i^\dagger\ket0,
    \qquad \ket{\tilde\Phi}=\prod_{i=1}^{r}\tilde\beta_i^\dagger\ket0.
\end{equation}
Writing $s_i=\cos\theta_i$, we obtain
\begin{equation}
    \label{eq:hatA}
    \begin{aligned}
        \ket{\tilde\Phi}&=e^{\hat A}\ket{\Phi},\\
        \hat A&=\sum_{i=1}^{t}\theta_i
        (\beta_{r+i}^\dagger\beta_i-\beta_i^\dagger\beta_{r+i})
        =\sum_{x,y=1}^{N}A_{xy}\beta_x^\dagger\beta_y.
    \end{aligned}
\end{equation}
The one-particle matrix $A$ is traceless and anti-Hermitian. It has nonzero entries only in the paired blocks, with $A_{r+i,i}=\theta_i$ and $A_{i,r+i}=-\theta_i$. Consequently,
\begin{equation}
    \label{eq:gutzAFrobenius}
    \|A\|_F^2=2\sum_{i=1}^{t}\theta_i^2.
\end{equation}

Lemma~\ref{lem:MainLemmaGutzwiller} in Appendix~\ref{sec:GutzwillerLemmas} gives
\begin{equation}
    \label{eq:gutzProjectedStability}
    \|\ket{\tilde\Psi}-\ket{\Psi}\|\leq mN\|A\|_F.
\end{equation}
Here the Frobenius norm is unchanged when $A$ is expressed in the original physical basis. We now relate this norm to the input error. The overlap is
\begin{equation}
    \label{eq:gutzOverlap}
    F:=|\langle\tilde\Phi|\Phi\rangle|
    =\prod_{i=1}^{r}s_i=\prod_{i=1}^{t}\cos\theta_i.
\end{equation}
Choosing the relative phase so that the overlap is nonnegative, Eq.~\eqref{eq:gutzInputApprox} implies $\epsilon^2\geq2(1-F)$. Since $\epsilon<1$, $F>0$, and $\cos\theta\leq e^{-\theta^2/2}$ gives
\begin{equation}
    \label{eq:gutzAngleBound}
    \sum_i\theta_i^2\leq-2\log F
    \leq-2\log(1-\epsilon^2/2)\leq2\epsilon^2.
\end{equation}
Thus $\|A\|_F\leq2\epsilon$, and
\begin{equation}
    \label{eq:gutzFinalError}
    \|\ket{\tilde\Psi}-\ket{\Psi}\|\leq2mN\epsilon.
\end{equation}
Given $\tilde\epsilon$, choose $\epsilon=\tilde\epsilon/(2mN)$. The local maps do not increase the virtual bond dimension beyond the product of the input dimensions, so
\begin{equation}
    \label{eq:gutzFinalBond}
    \tilde D\leq D^m=\operatorname{poly}(N,1/\tilde\epsilon)
\end{equation}
for fixed $m$. The empty and fully occupied cases are immediate.
\end{proof}

For the Chern insulator of Sec.~\ref{sec:gappedGaussianModels}, $N=n^2$ and $r=n^2/2$. Taking $m=2$ gives the projected chiral state discussed in Sec.~\ref{sec:gutzwillerModels}, with bond dimension polynomial in $n$ and $1/\tilde\epsilon$. Theorem~\ref{thm:gappedNumberConserving} also supplies Gaussian input approximations for other uniformly gapped, finite-range Chern models. 

% ===================================================================
\section{Summary and Outlook}

We have established efficient PEPS approximations for critical
free-fermion states, gapped Gaussian states, and a family of chiral
topological states. Our results include a specified half-filled ground
state of the two-dimensional nearest-neighbor tight-binding model and
the ground states of uniformly gapped, finite-range quadratic
Hamiltonians in any fixed spatial dimension, including Chern insulators
and gapped $p+ip$ superconductors. We have also shown that
equal-occupancy Gutzwiller projections of a fixed number of identical
Slater determinants at arbitrary filling admit efficient PEPS
approximations whenever the input states admit efficient Gaussian PEPS
approximations. This includes the lattice Laughlin-type and
chiral-spin-liquid states considered here. In all these cases, the
global norm error can be made at most $\epsilon$ with bond dimension
$D=\operatorname{poly}(n,1/\epsilon)$.

Several directions remain to be explored. It would be interesting to
extend these results to more general chiral topological states, beyond
the projected families considered here, and to interacting critical
fermionic models whose entanglement entropy obeys an area law with
multiplicative logarithmic corrections, such as
$S(\ell)=O(\ell^{d-1}\log\ell)$ for regions of linear size $\ell$ in
$d$ dimensions. More generally, an important question is whether this
entanglement scaling, together with translation invariance, is
sufficient to guarantee PEPS approximations with polynomial bond
dimension, or whether additional assumptions are necessary. Another
natural direction is to extend the results to lattice gauge theories,
while respecting their local gauge constraints, and to investigate
the continuum limit. The latter would require controlling the
dependence of the bond dimension and approximation error on the
lattice spacing, in addition to the system size.

% ===================================================================
\acknowledgments

We thank Subir Sachdev for raising the question of whether free fermion models with a Fermi surface can be effciently approximated by PEPS. RT acknowledges useful discussions with Georgios Styliaris. The authors acknowledge minor use of AI models (GPT 5.6 Sol, GPT 6.1 Astra) for assisting with this paper. In particular, AI was used to check the proofs and look for references, as well as for editing the text. This work is partly funded by the Munich Quantum Valley, which is supported by the Bavarian state government with funds from the Hightech Agenda Bayern Plus, and by the Munich Center for Quantum Science and Technology (MCQST), funded by the Deutsche
Forschungsgemeinschaft (DFG) under Germany’s Excellence Strategy (EXC2111-390814868). R.T. acknowledges support from the European Union’s Horizon Europe research and innovation program under grant agreement number 101221560 (ToNQS).

% ===================================================================
\appendix
\section{Approximation of the 1D ground state by an MPS}
\label{sec:appendixMPS}

In this section, we prove Lemma~\ref{lem:chainTail}. We want now to upper-bound the rhs of
Eq.~\eqref{eq:truncationError}. At each cut, we define
\begin{equation}
    \label{eq:display014}
    Z_a= \sum_{j} \sqrt{\lambda_j}
\end{equation}
Since the eigenvalues are sorted in decreasing order, $D\sqrt{\lambda_{D+1}}\leq Z_a$. Thus,
\begin{equation}
    \label{eq:schmidtTailEstimate}
    t_a(D) \leq \sqrt{\lambda_{D+1}} \sum_{j>D}\sqrt{\lambda_j} \leq \frac{Z_a^2}{D}
\end{equation}
If fewer than $D+1$ eigenvalues are nonzero, this bound holds trivially.

We next bound $Z_a$ uniformly in the cut $a$. Because the ground state $\ket{\omega}$ is a Slater determinant~\cite{botero2004,schuch2019}, we can choose fermionic modes $f_1,\ldots,f_{2a}$ supported in $A_a$ such that the
reduced density operator factorizes:
\begin{equation}
    \label{eq:display015}
    \rho_A =\bigotimes_{\alpha=1}^{2a}
    \left[(1-\nu_\alpha)\ket{0_\alpha}\bra{0_\alpha}+\nu_\alpha\ket{1_\alpha}\bra{1_\alpha}\right].
\end{equation}
Here $\nu_\alpha$ is the occupation probability of mode $f_\alpha$. The eigenvalues of $\rho_A$, which
are the squared Schmidt coefficients, are therefore
\begin{equation}
    \label{eq:reducedSpectrum}
    \lambda_{\boldsymbol s} = \prod_{\alpha=1}^{2a} \nu_\alpha^{s_\alpha}(1-\nu_\alpha)^{1-s_\alpha},
    \qquad \boldsymbol s\in\{0,1\}^{2a}.
\end{equation}
It follows immediately that
\begin{equation}
    \label{eq:logZ}
    \log Z_a = \frac12\sum_{\alpha=1}^{2a} \log\left(1+2\sqrt{\nu_\alpha(1-\nu_\alpha)}\right).
\end{equation}

Expression~\eqref{eq:logZ} can be directly connected to the first quantization description of the
problem. Indeed, let us denote by $\ket{c_x}$ and  $\ket{d_x}$ the single-particle states located
at position $x$ and in the $c$ and $d$ orbitals. For a region $A$, we define
\begin{equation}
    \label{eq:display016}
    P_A = \sum_{x\in A} \ket{c_x}\bra{c_x} + \ket{d_x}\bra{d_x}, \quad P_{A^c}=\Id-P_A
\end{equation}
Let us further define as $P$ the projection onto the 1-particle states that are occupied in the
ground state. The operator $P_A P P_A$ has eigenvalues $\nu_\alpha$. Defining $X= P_A P (\Id-P_A)$,
since $P^2=P$ we immediately deduce that $X$ has singular values precisely
$\sqrt{\nu_\alpha(1-\nu_\alpha)}$, so that we can bound $\log Z_a \le \|X\|_1$.

In our problem, the occupied orbitals have the form
\begin{equation}
    \label{eq:display017}
    \ket{\eta_y} = \frac{1}{\sqrt{2}} \left( \sum_{x=0}^{n-1} G_{x,y}\ket{c_x} + \ket{d_y}\right).
\end{equation}
Here $G$ is given in Eq.~\eqref{eq:matrixG}, and can be explicitly computed using
Eq.~\eqref{eq:groundState1D},
\begin{equation}
    \label{eq:display018}
    G_{x,y} = \frac{1}{n} \left[ 1+\cot\left(\frac{\pi r_{xy}}{n}\right)\right],
\end{equation}
with $r_{xy}=y-x+1/2$. Since $G$ is unitary, $\|X\|_1= \|Q_A G Q_{A^c}\|_1= \|Q_{A^c} G Q_A\|_1$,
where $Q_A$ is the projector on the lattice sites in region $A$. To bound this quantity, we define
the rectangular matrices
\begin{equation}
    \label{eq:display019}
    C_{x,y} = \frac{1}{n}\cot\left(\frac{\pi r_{xy}}{n}\right), \qquad S_{x,y}
    = \frac{1}{n\sin(\pi r_{xy}/n)},
\end{equation}
where $x\in A,\quad y\in A^c$. Then $C=(T+\overline{T})/2$, where $T_{x,y}= e^{-i\pi
x/n}e^{i\pi(y+\frac12)/n}S_{x,y}$. Consequently, $\|C\|_1\leq\|T\|_1=\|S\|_1$. The constant block
has rank one and trace norm $\sqrt{|A||A^c|}/n\leq 1/2$, so
\begin{equation}
    \|X\|_1  \leq \frac12+\|S\|_1.\label{eq:nuclearNormBound}
\end{equation}

In order to bound $\|S\|_1$, we use the integral representation
\begin{equation}
\label{eq:integralS}
    S_{xy}=\frac1\pi\int_0^\infty \frac{e^{-r_{xy}t}+e^{-(n-r_{xy})t}}{1+e^{-nt}}\,dt.
\end{equation}
The two exponentials in the integrand are rank-one matrices which we denote as $E(t)$ and $F(t)$,
respectively. Since $x=0,\ldots,a-1$, $y=a,\ldots,n-1$ we can write the first exponential as
\begin{equation}
    \label{eq:display020}
    E_{xy}(t) =\frac{e^{-3t/2}}{1+e^{-nt}}\, e^{-(a-1-x)t}e^{-(y-a)t}.
\end{equation}
so that the trace-norm for $t>0$ is
\begin{equation}
    \label{eq:display021}
    \|E(t)\|_1 = \frac{e^{-3t/2} \sqrt{(1-e^{-2at})(1-e^{-2(n-a)t})}} {(1+e^{-nt})(1-e^{-2t})}.
\end{equation}
The numerator can be bounded by $1-e^{-nt}$, which gives $\|E(t)\|_1 \le e^{-t/2}
\tanh(nt/2)/2\sinh(t)$. An analogous calculation shows that $\|F(t)\|_1\leq e^{t/2}
\tanh(nt/2)/2\sinh(t)$, so that
\begin{equation}
    \label{eq:display022}
    \|S\|_1 \leq\frac1\pi\int_0^\infty \frac{\tanh(nt/2)}{2\sinh(t/2)}\,dt.
\end{equation}
Using $\tanh u\leq\min\{u,1\}$ and $2\sinh(t/2)\geq t$, and splitting the integral at $1/n$ and
$1$, yields
\begin{equation}
    \label{eq:display023}
        \|S\|_1 \leq \frac1\pi \left[\frac12+\log n+\log\coth\left(\frac14\right)\right]\leq\frac{\log n+2}{\pi}.
\end{equation}
Combining this with Eq.~\eqref{eq:nuclearNormBound} and $\log Z_a\leq\|X\|_1$ gives, uniformly at every
cut,
\begin{equation}
    \label{eq:display024}
    \log Z_a \leq\frac12+\frac{\log n+2}{\pi}
\end{equation}
Consequently,
\begin{equation}
    \label{eq:mpsRenyiBound}
    Z_a^2\leq C_0n^{2/\pi},\qquad C_0=e^{1+4/\pi}.
\end{equation}
Together with Eq.~\eqref{eq:schmidtTailEstimate}, this proves Lemma~\ref{lem:chainTail}.

% ======================================================
\section{Approximation of the wavelet operator as an MPU}
\label{sec:appendixMPU}

In this Appendix we prove the results presented in Sec.~\ref{sec:waveletMPU}. We will
consider a lattice with $n=2m$ sites and write subindices whenever it is needed in order to clarify
to which lattice size we refer to. For instance, we will use $\mathcal X_n$ to denote the set of $n$ lattice
sites, or $G_n$ the matrix $G$ of Eq.~\eqref{eq:matrixG} in such a lattice.

% ------------------------------------------------------
\subsection{Construction of $H$ and $J$} \label{sec:filterConstruction}

Here we will give explicit unitary matrices $H$ and $J$ fulfilling Eq.~\eqref{eq:waveletIdentity}, and which are
used for the construction of the wavelet transformation. Let $n=2m$. We define filter functions
\begin{equation}
    \label{eq:filterDefinitions}
    h(k)= \frac{\sqrt{2}\cos^{2p}(k/2)} {\cos^{2p}(k/2)+i\sin^{2p}(k/2)},\qquad j(k)=e^{-ik/2}h(k)
\end{equation}
where $p$ is a parameter which will be chosen later. In the limit $p\to\infty$, $|h|$ and $|j|$
tend to $\sqrt2$ for $|k|<\pi/2$ and to zero for $\pi/2<|k|\leq\pi$; at $|k|=\pi/2$ their magnitude
is one. Thus they can be considered as filter functions with finite width. One can readily check
that
\begin{equation}
    \label{eq:filterPhaseIdentities}
    \gamma(k)j(k)=h(k)\gamma(2k), \qquad \gamma(k)\overline{j(k+\pi)} =\overline{h(k+\pi)}
\end{equation}
which will be needed later on.

Following the wavelet construction of Refs.~\cite{evenbly2016wavelets,haegeman2018wavelets}, we now build the matrices $H$ and $J$ fulfilling Eq.~\eqref{eq:waveletIdentity} as follows
\begin{subequations}
    \label{eq:filterBankColumns}
    \begin{align}
    \label{eq:display025}
        H_{x,y}&=h_o(x-2y),\qquad     H_{x,m+y}=h_e(x-2y),\\
        J_{x,y}&=j_o(x-2y),\qquad     J_{x,m+y}=j_e(x-2y),
     \end{align}
\end{subequations}
where $0\leq x<n$ and $0\leq y<m$, and where for a function $f(k)$ we define
\begin{subequations}
    \label{eq:filterProfiles}
    \begin{align}
    \label{eq:display026}
        f_o(x) &=\frac{1}{n}\sum_{k\in\mathcal K_n}f(k)e^{ikx},\\
        f_e(x) &= \frac{1}{n}\sum_{k\in\mathcal K_n} \overline{f(k+\pi)}\,e^{ik(x-1)}.
    \end{align}
\end{subequations}
Note that both matrices, $H$ and $J$ are unitary. Indeed, one can easily check that their
corresponding columns are orthonormal just by noticing that
\begin{equation}
    \label{eq:aliasNormalization}
    |f(k)|^2+|f(k+\pi)|^2=2
\end{equation}
for both $f=h,j$ and for all $k\in \mathcal K_n$.

Now we will show that they fulfill Eq.~\eqref{eq:waveletIdentity}. Substituting the Fourier sums gives
\begin{equation}
\label{eq:filterFirstIdentity}
    \sum_{z=0}^{n-1}(G_n)_{x,z}\,j_o(z-2y) =\sum_{v=0}^{m-1}h_o(x-2v)(G_m)_{v,y},
\end{equation}
This can be seen by noticing that
\begin{align}
    \label{eq:display027}
    \sum_{z=0}^{n-1}(G_n)_{x,z}j_o(z-2y)  &= \frac{1}{n}\sum_{k\in\mathcal K}
    \gamma(k)j(k)e^{ik(x-2y)}\\ \nonumber
    \sum_{v=0}^{m-1}h_o(x-2v)(G_m)_{v,y} &=  \frac{1}{n}\sum_{k\in\mathcal K}
    h(k)\gamma(2k)e^{ik(x-2y)}.
\end{align}
and Eq.~\eqref{eq:filterPhaseIdentities}. We also have
\begin{equation}
    \label{eq:filterSecondIdentity}
    \sum_{z=0}^{n-1}(G_n)_{x,z}\,j_e(z-2y) =h_e(x-2y).
\end{equation}
which follows from Eq.~\eqref{eq:filterPhaseIdentities}.

Writing $H=(H_o\;\;H_e)$ and $J=(J_o\;\;J_e)$, Eqs.~\eqref{eq:filterFirstIdentity} and \eqref{eq:filterSecondIdentity} state that
\begin{equation}
    \label{eq:display028}
    G_nJ=(H_oG_m\;\;H_e)
    =H(G_m\oplus \Id_m).
\end{equation}
Since $H$ is unitary, we obtain Eq.~\eqref{eq:waveletIdentity}.

% ------------------------------------------------------
\subsection{Bounds on the MPO bond dimension}
\label{sec:mpoBounds}

As we mentioned in Sec.~\ref{sec:waveletMPU}, the wavelet transformation $U$ can be split
into the product of two commuting operators, $U_n^c$ and $U_n^d$. They both act by multiplying the
corresponding mode operator $\boldsymbol c$ and $\boldsymbol d$ by a certain unitary matrix (see
Eq.~\eqref{eq:waveletAction}). In this subsection we will show that if $C$ can be expressed in terms of
some polynomials of degree $p$, then we can write the corresponding unitary by an MPU with bond
dimension that is bounded exponentially in $p$. Later on we will show that, indeed, $C_h$ fulfills
that property, which automatically provides such a bound for the MPU describing $U^c$. For $C_j$ this
is not the case, and this is why in the next subsection we will have to approximate it by another
matrix fulfilling that property, which provides an approximation to $U^d$ with the same bond
dimension.

Let us start considering a function $f(k)$ and a matrix $C$ built as
\begin{subequations}
    \label{eq:filterColumns}
    \begin{align}
    \label{eq:display029}
        (C)_{x,2t}&=\frac1n\sum_{k\in\mathcal K_n}f(k)e^{ik(x-2t)}\\
        (C)_{x,2t+1}&=\frac1n\sum_{k\in\mathcal K}
        \overline{f(k+\pi)}e^{ik(x-2t-1)},
    \end{align}
\end{subequations}
where $x\in\mathcal X_n$, $t\in\mathcal X_m$. If we choose $f=h,j$ then we obtain the corresponding
$C_{h,j}$ defined above (see Eqs.~\eqref{eq:waveletMatrices} and \eqref{eq:filterProfiles}). Note that the matrix $Q$ used
in that definition implies that the columns of $C_{h,j}$ are reshuffled as compared to those of $H$
and $J$.

Now, for a matrix $C$ we define the unitary operator $U_C$ that acts as in
Eq.~\eqref{eq:waveletAction}. That is, if we denote as before by $\boldsymbol a$ a column vector of
annihilation operators,
\begin{equation}
    \label{eq:display030}
    U_C \boldsymbol a^\dagger U_C^\dagger = \boldsymbol a^\dagger C,\qquad U_C\ket0=\ket0.
\end{equation}
This operator is Gaussian and can be written as an MPO. What we want to do now is to bound its bond
dimension. For that we consider, as before, a set of sites $A=(0,\ldots,a-1)$ and its complement
$A^c$. We will bound the Schmidt rank of this operator with respect to those two regions, which
will then give us a bound to the bond dimension, which we will call $D_A$.

First, we will show that the bond dimension is related to the rank of some matrix connected to $C$.
More specifically, let us define $Q_A=\sum_{x\in A}\ket{x}\bra{x}$ and $Q_{A^c}=\Id-Q_A$,
consistently with the site projectors used in Sec.~\ref{sec:appendixMPS}. We also define
\begin{equation}
    \label{eq:crossRank}
    r_A=\operatorname{rank} C_{A,A^c}  =\operatorname{rank}(Q_A C Q_{A^c}).
\end{equation}
Let us show that we can relate the bond dimension $D_A$ of the MPU describing exactly $U_C$ to this
rank:

\begin{lemma}
    \label{lem:operatorSchmidtRank}
    Let $C$ be unitary. Then $D_A\leq4^{r_A}$.
\end{lemma}
\begin{proof}
The two off-diagonal blocks of a unitary matrix have equal rank $r=r_A$. Choose unitaries $W_A$ and
$W_{A^c}$ whose last $a-r$ and $n-a-r$ columns span the kernels of $C_{A^c,A}$ and $C_{A,A^c}$,
respectively. Then $\tilde C=C(W_A\oplus W_{A^c})$ has only $r$ nonzero columns in each
off-diagonal block. The corresponding $U_{\tilde C}$ differs from $U_C$ by a product of local
unitaries, so its operator Schmidt rank is unchanged. We omit the tilde below.

We use the fermionic Choi state~\cite{bravyi2005}, assigning each auxiliary mode $b_t$ to the same
spatial region as $a_t$:
\begin{equation}
    \label{eq:gaussianChoiState}
    \ket{\Psi_C}=\prod_{t\in\mathcal X_n}g_t^\dagger\ket0,
\end{equation}
where
\begin{equation}
    \label{eq:display031}
    g_t^\dagger=\frac{\sum_{x\in\mathcal X_n}C_{x,t}a_x^\dagger+b_t^\dagger}{\sqrt2}.
\end{equation}
The Schmidt rank of this state, with both copies of each region kept together, is $D_A$. All but
$r$ of the factors with $t\in A$ are supported entirely in $A$, and all but $r$ of the factors with
$t\in A^c$ entirely in $A^c$. Each remaining factor is a sum of two operators, one on either side.
Expanding these at most $2r$ factors gives at most $2^{2r}=4^r$ product vectors. Fermionic
reordering changes only their signs, proving the claim.
\end{proof}

Next, we will show that if $C$ is defined as in Eq.~\eqref{eq:filterColumns} and $f(k)$ is the
quotient of polynomials in $z=e^{ik}$, then $r_A$ is bounded by the degree of such polynomials.
Note that for $C_h$, such a degree is at most $2p$ (Eq.~\eqref{eq:filterDefinitions}), so this, together with the
previous lemma, will provide a useful upper bound for $U^c$. As we mentioned above, this does not
apply to $C_j$, since the factor $e^{-ik/2}$ cannot be written as quotients of polynomials. In that
case we will need a further approximation.

In order to proceed, we define another function $g(z)=f(k)$, $\tilde g(z)=\overline{f(k+\pi)}$ for
$z=e^{ik}$. This means, e.g., that we write $f(k)$ in terms of $e^{i k m}$ and then replace those
terms by $z^m$.

\begin{lemma}
    \label{lem:filterRank}
    Let $C_f$ be defined by Eq.~\eqref{eq:filterColumns} with $g(z)=p_0(z)/p_1(z)$,
    with $\deg p_0,\deg p_1\leq d$, and $p_1$ has no zeros in $|z|=1$. Then, for any
    contiguous interval $A$, $r_A\leq2d$.
\end{lemma}

\begin{proof}
Let us define the shift operator
\begin{equation}
    \label{eq:display032}
    S=\sum_{t\in\mathcal X_n}\ket{t-1}\bra{t}
\end{equation}
One can readily see that
\begin{equation}
    \label{eq:filterShiftMatrix}
        (C)_{x,2t}=\left[\ g(S)\right]_{x,2t},\quad  (C)_{x,2t+1}=\left[\ \tilde g(S)\right]_{x,2t+1}.
\end{equation}
So that we can write
\begin{equation}
    \label{eq:filterEvenOdd}
    C=g(S)E+\tilde g(S)O,
\end{equation}
where
\begin{equation}
    \label{eq:display033}
     E=\sum_{t\in\mathcal X_m}\ket{2t}\bra{2t},\qquad O=\Id-E
\end{equation}

Let us denote $r_A(M)=\operatorname{rank}(Q_A M Q_{A^c})$. Note that $r_A(XE)\le r_A(X)$ since $E$ just suppresses
some columns of $X$. The same is true for $r_A(YO)$. Furthermore, $r_A(X+Y)\le r_A(X)+r_A(Y)$.
Thus, according to Eq.~\eqref{eq:filterEvenOdd},
\begin{equation}
    \label{eq:display034}
    r_A=r_A(C)\le r_A(g(S))+r_A(\tilde g(S)).
\end{equation}
We will show that if $g(z)$ is a quotient of polynomials of degree at most $d$, then $r_A(g(S))\le
d$. Note that for $|z|=1$, $\tilde g(z)=\overline{g(-z)}$, so that $\tilde g(z)$ is also a quotient
of polynomials of degrees at most $d$, from which the proof of the lemma will follow.

Let us set $T=p_1(S)^{-1}p_0(S)=g(S)$ and $A=\{0,\ldots,a-1\}$. The case $d\geq a$ is immediate.
Otherwise put $I=\{0,\ldots,a-d-1\}$. Row $x\in I$ of a polynomial in $S$ of degree at most $d$ has
support only at $x,\ldots,x+d$, all in $A$. Hence
\begin{equation}
    \label{eq:display035}
    [p_0(S)]_{I,A^c}=[p_1(S)]_{I,A^c}=0.
\end{equation}
Taking the $I,A^c$ block of $p_1(S)T=p_0(S)$ gives
\begin{equation}
    \label{eq:filterKernelConstraint}
    [p_1(S)]_{I,A}\,T_{A,A^c}=0.
\end{equation}
The matrix $p_1(S)$ is invertible, and its rows indexed by $I$ are independent. They vanish outside
$A$, so $[p_1(S)]_{I,A}$ has rank $a-d$ and a kernel of dimension $d$. Every column of $T_{A,A^c}$
belongs to this kernel, proving $r_A(g(S))\leq d$. The same bound applies to $\tilde g(S)$, and
therefore $r_A(C)\leq2d$. A cyclic translation gives the same result for any contiguous interval.

\end{proof}

% ------------------------------------------------------
\subsection{Rational approximation of the Gaussian transformation}
\label{sec:rationalApproximation}

We now apply Lemmas~\ref{lem:operatorSchmidtRank} and \ref{lem:filterRank} to the operator $U=U^cU^d$
in Eq.~\eqref{eq:waveletAction}. We can consider $U^c$ and $U^d$ separately, and simply multiply
the corresponding upper bounds to their bond dimension, $D_c$ and $D_d$, respectively, in order to
obtain the bond dimension, $D$, of the MPU corresponding to $U$.

Let us start with the simplest case of $U^c$. There we can directly use the results derived in the
previous subsection to upper bound the bond dimension.
\begin{lemma}\label{lem:cModeMPO}
    $U^c$ can be written exactly as an MPU with bond dimension $D_c\leq2^{16p}$.
\end{lemma}

\begin{proof}
    The filter $h(k)$ is already rational in $z=e^{ik}$:
\begin{equation}
    \label{eq:rationalH}
    h(k)=\frac{\sqrt2(z+1)^{2p}}
    {(z+1)^{2p}+i(-1)^p(z-1)^{2p}}.
\end{equation}
Both polynomials have degree at most $2p$, and the denominator has no zeros on the unit circle.
Thus, Lemmas~\ref{lem:operatorSchmidtRank} and \ref{lem:filterRank} bound the bond dimension. Additionally, since the description
is exact, the corresponding MPO is unitary, and thus it is an MPU.
\end{proof}

The filter $j(k)=e^{-ik/2}h(k)$ defining $U_n^d$ cannot be written as quotient of polynomials in
$z$. In order to circumvent this obstacle, we approximate $e^{-ik/2}$ by a function $r(k)$ which is
a quotient of polynomials in $z$. This function will depend on the parameter $p$, which
adjusts the degree of the polynomials as well as the degree of approximation. Thus, replacing
$j(k)$ by $r(k)h(k)$ will allow us to approximate $U^d$ by an MPO with a suitable bond dimension.
Additionally, we will make sure that the approximation is unitary.

For $p\geq1$, set $s=e^{ik/2}$ and define
\begin{equation}
    \label{eq:rationalHalfShift}
    r(k)=\frac{(1+s)^{2p+1}-(1-s)^{2p+1}}
    {s\left[(1+s)^{2p+1}+(1-s)^{2p+1}\right]}.
\end{equation}
This is a quotient of two degree-$p$ polynomials in $z=s^2=e^{ik}$. Indeed, the sum
$(1+s)^{2p+1}+(1-s)^{2p+1}$ is an even polynomial of degree $2p$ in $s$. The difference is an odd
polynomial of degree $2p+1$; dividing it by $s$ gives another even polynomial of degree $2p$. Both
are therefore polynomials of degree $p$ in $z=s^2$.

It is convenient to rewrite
\begin{equation}
    \label{eq:rationalPhase}
    r(k)=e^{-ik/2}\frac{1-w^{2p+1}}{1+w^{2p+1}},\qquad w=\frac{1-s}{1+s}=-i\tan(k/4).
\end{equation}
For real $k$, the number $w^{2p+1}$ is purely imaginary, so the numerator and denominator of the
second factor are complex conjugates. Thus, $|r(k)|=1$. There are also no poles on the unit circle,
and therefore it is suited for the application of Lemma~\ref{lem:filterRank}.

We further define
\begin{equation}
    \label{eq:rationalJ}
    \tilde j(k)=r(k)h(k).
\end{equation}
This allows us to introduce $\tilde J$ as in Sec.~\ref{sec:filterConstruction}, and
$\tilde U^d$ as in Eq.~\eqref{eq:waveletDAction}. Then, we have the following

\begin{lemma}\label{lem:dModeMPO}
    $\tilde U^d$ is exactly unitary and can be written exactly as an MPU with bond dimension $D_d\leq2^{24p}$ and
    \begin{equation}
    \label{eq:display036}
        \|U^d-\tilde U^d\|\le 2\sqrt{2} n \rho^p
    \end{equation}
    where $\rho=3-2\sqrt2$
\end{lemma}

\begin{proof}
The fact that the bond dimension is bounded follows directly from Lemmas~\ref{lem:operatorSchmidtRank}
and \ref{lem:filterRank}. Furthermore, notice that $\tilde J$ is unitary. This follows from the
fact that $|\tilde j(k)|=|h(k)|$, and therefore
\begin{equation}
    \label{eq:rationalJNormalization}
    |\tilde j(k)|^2+|\tilde j(k+\pi)|^2=2.
\end{equation}
which, as mentioned in Sec.~\ref{sec:filterConstruction} implies unitarity. Thus, $\tilde
U^d$ is also unitary, and an MPO description will yield an MPU. It only remains to derive how much
it departs from $U^d$.

Defining $u=|\tan(k/4)|$, direct subtraction gives
\begin{equation}
    \label{eq:halfShiftError}
    \left|r(k)-e^{-ik/2}\right| =\frac{2u^{2p+1}}{\sqrt{1+u^{4p+2}}}
    \leq2u^{2p+1}.
\end{equation}
Similarly, setting $v=|\tan(k/2)|$, Eq.~\eqref{eq:filterDefinitions} gives
\begin{equation}
    \label{eq:hMagnitudeBound}
    |h(k)|=\frac{\sqrt2}{\sqrt{1+v^{4p}}} \leq\sqrt2\min\{1,v^{-2p}\}.
\end{equation}
Using Eq.~\eqref{eq:halfShiftError} and setting $v=2u/(1-u^2)$, we obtain
\begin{equation}
    \label{eq:weightedHalfShiftError}
        |\tilde j(k)-j(k)|
        \le 2\sqrt2\,u\left[\min\left\{u,\frac{1-u^2}{2}\right\}\right]^{2p}.
\end{equation}
For $0\leq u\leq1$, the minimum in this expression is at most $\sqrt2-1$. The first function
increases, the second decreases, and they intersect at $u=\sqrt2-1$. Consequently,
\begin{equation}
    \label{eq:uniformJError}
        |\tilde j(k)-j(k)|
        \leq2\sqrt2\,\rho^p = \delta,\qquad  \rho=(\sqrt2-1)^2=3-2\sqrt2.
\end{equation}
uniform in $k$.

In momentum representation $\tilde J-J$ is block diagonal, so that the operator norm is the maximum
of the operator norms of those blocks, which are bounded by $\delta$, yielding that
$\|\tilde J-J\|\le \delta$. This immediately implies that (cf Eq.~\eqref{eq:waveletIdentity})
\begin{equation}
    \label{eq:approximateWaveletIdentity}
    \left\|H^\dagger G_n\tilde J-(G_m\oplus\Id_m)\right\|_{\mathrm{op}}
    \leq2\sqrt2\,\rho^p.
\end{equation}
From that we obtain the desired result
\begin{equation}
    \label{eq:rationalUnitaryError}
    \begin{aligned}
        \|\tilde U_n^d-U_n^d\|_{\mathrm{op}}
        \leq n\|\tilde J-J\|_{\mathrm{op}}
        \leq2\sqrt2\,n\rho^p.
    \end{aligned}
\end{equation}
where we have used that the distance between second quantized unitaries in a sector of $r$
particles is bounded by $r$ times the distance between the first quantized unitaries. Furthermore,
multiplication by $Q$ and taking adjoints leave the one-particle norm unchanged.

Finally, by Lemmas~\ref{lem:operatorSchmidtRank} and \ref{lem:filterRank}, we obtain an exact MPO
representation of $\tilde U^d$ with $D_d\leq2^{24p}$.

\end{proof}

The bonds of the two species multiply, while their local physical indices are combined into the
four-dimensional space of one $(c_x,d_x)$ cell. For a prescribed one-step error $0<\epsilon<1$,
choose
\begin{equation}
    \label{eq:rationalFilterChoice}
    p=\left\lceil\frac{\log(2\sqrt2\,n/\epsilon)}{\log(1/\rho)}\right\rceil.
\end{equation}
Then
\begin{equation}
    \label{eq:rationalUnitaryConclusion}
    \|\tilde U_n-U_n\|_{\mathrm{op}}\leq\epsilon,
    \qquad D(\tilde U_n)=\operatorname{poly}(n,1/\epsilon).
\end{equation}
The Gaussian replacement is exactly unitary and its MPO is exact. Only $U_n^d$ is approximated;
$U_n^c$ is unchanged. These single-step bounds do not by themselves bound the complete
one-dimensional product $\mathcal U$. The two-dimensional arrangement is treated in the PEPS
construction.

% ======================================================
\section{PEPS construction}
\label{sec:appendixPEPS}

In this Appendix we give the precise way of performing the fermionic swap operators required to
construct the PEPS, and also compute the bond dimension and the error.

% ------------------------------------------------------
\subsection{Fermionic swap operations} \label{sec:fermionicSWAPs}

For two neighboring cells $\boldsymbol u,\boldsymbol v$, we denote by $F_{\boldsymbol u,\boldsymbol
v}$ the fermionic SWAP that exchanges both species. Thus
\begin{equation}
    \label{eq:cellSWAPAction}
    F_{\boldsymbol u,\boldsymbol v}\,\sigma_{\boldsymbol u}^\dagger
    F_{\boldsymbol u,\boldsymbol v}^\dagger=\sigma_{\boldsymbol v}^\dagger,
    \qquad\sigma=c,d,
\end{equation}
This is a Gaussian operation which leaves the vacuum unchanged. For the first exchange, we take
$\boldsymbol u=(4r,4s)$ and apply, in chronological order,
\begin{equation}
    \label{eq:firstCellExchange}
    F_{\boldsymbol u,\boldsymbol u+\boldsymbol e_x},\quad
    F_{\boldsymbol u+\boldsymbol e_x,\boldsymbol u+\boldsymbol e_x+\boldsymbol e_y},\quad
    F_{\boldsymbol u,\boldsymbol u+\boldsymbol e_x}.
\end{equation}
This exchanges $\boldsymbol u$ and $\boldsymbol u+\boldsymbol e_x+\boldsymbol e_y$ and restores the
intermediate cell, even if that cell belongs to an entangled chain. For example, the complete cell
labels at $(0,0),(1,0),(1,1)$ evolve as
\begin{equation}
    \label{eq:display037}
    (A,B,D)\longrightarrow(B,A,D)\longrightarrow(B,D,A)
                 \longrightarrow(D,B,A).
\end{equation}
The product of these completed exchanges, for $r,s\in\mathcal X_{n/4}$, defines $P_1$.

At every later step we do the same. Write $W_j$ the wavelet unitary acting on the active cells
$(x_j(t),x_j(u))$, and as the identity elsewhere. After $W_j$, the four groups in
Eq.~\eqref{eq:sectorPositions2D} are at
\begin{equation}
    \label{eq:display038}
    \bigl(x_j(2r+\mu),x_j(2s+\nu)\bigr),
    \quad \mu,\nu\in\{0,1\},\quad r,s\in\mathcal X_{n_j/2}.
\end{equation}
The cells of the smaller two-dimensional state, and their destinations, are
\begin{subequations}
    \label{eq:swapEndpoints}
    \begin{align}
    \label{eq:display039}
        \boldsymbol u_{r,s}^{(j)}
        &=\bigl(2^{j+1}r+2^{j-1}-1,\,2^{j+1}s+2^{j-1}-1\bigr),\\
        \boldsymbol v_{r,s}^{(j)}
        &=\boldsymbol u_{r,s}^{(j)}+(\delta_j,\delta_j),
          \qquad\delta_j=2^{j-1}.
    \end{align}
\end{subequations}
In particular, $\boldsymbol v_{r,s}^{(j)}=(x_{j+1}(r),x_{j+1}(s))$. We thus define $P_j$ to
exchange each complete cell at $\boldsymbol u_{r,s}^{(j)}$ with the one at $\boldsymbol
v_{r,s}^{(j)}$.

Every destination still contains one of the step-zero dimers in the exact construction. Both of its
coordinates are odd, and it belongs to $\mathcal X^{(j+1)}\times\mathcal X^{(j+1)}$, disjoint from
the supports of the earlier Gaussian steps and the endpoints of earlier exchanges. Furthermore,
intermediate cells touched during an exchange are restored before the next operation. Thus $P_j$
moves the smaller state to its next active lattice and moves the destination dimers to the vacated
cells. The released critical chains remain at their release positions.

For $\boldsymbol u=(x,y)$ and $\boldsymbol v=(x+\delta_j,y+\delta_j)$, we implement the exchange
along the nearest-neighbor path
\begin{equation}
    \label{eq:swapPath}
    (x,y)\longrightarrow(x+\delta_j,y)
         \longrightarrow(x+\delta_j,y+\delta_j).
\end{equation}
If $F_1,\ldots,F_{L_j}$ are its successive adjacent-cell SWAPs, with $L_j=2\delta_j=2^j$, the
chronological sequence is
\begin{equation}
    \label{eq:swapSequence}
    F_1,F_2,\ldots,F_{L_j},F_{L_j-1},\ldots,F_2,F_1.
\end{equation}
This procedure uses $2L_j-1$ fermionic SWAPs.

% ------------------------------------------------------
\subsection{Proof of the Lemmas} \label{sec:appendixPEPSProofs}

In this subsection we prove Lemmas~\ref{lem:pepsBond} and \ref{lem:pepsError}.

\begin{proof}[Proof of Lemma~\ref{lem:pepsBond}]
By Lemmas~\ref{lem:cModeMPO} and \ref{lem:dModeMPO}, a complete one-dimensional operation on both species has
an exact MPO bond at most $D_U=2^{40p}$. At a coarse scale, we extend the bond between consecutive
active sites through the intermediate sites by identity tensors. This leaves its dimension
unchanged. Since a fixed row is in at most one $\mathcal X^{(j)}$ for $j\geq1$, a horizontal edge
carries at most one wavelet row-MPO bond and one bond from $W_0$. The same statement holds
vertically.

For the SWAPs, the horizontal part of a step-$j$ path lies on a row in $\mathcal X^{(j)}$, and its
vertical part on a column in $\mathcal X^{(j+1)}$. These sets are disjoint across steps. Within one
step, adjacent sources along a row are separated by $2^{j+1}=4\delta_j$, while the horizontal
segment has length $\delta_j$; the corresponding vertical statement is identical. Hence no two
distinct paths use the same edge and Eq.~\eqref{eq:swapSequence} uses each path edge at most
twice. A neighboring-cell fermionic SWAP has operator Schmidt rank at most $4^2=16$, because each
cell has local dimension four. Thus all permutations contribute at most $16^2$ per edge.

The step-zero horizontal chains lie on odd rows. A later horizontal chain lies on a row
$x_j(2s+1)\in\mathcal X^{(j)}$, and no later permutation moves its cells. Therefore a row contains
chains from at most step zero and one later release. Therefore, open MPS bonds connecting their
consecutive cells give at most two bonds of dimension $D_1$ on any horizontal edge. The vertical
count is the same.

We then contract the MPO tensors and SWAP tensors in their prescribed temporal order, and fuse all
tensors at each physical cell. Local intermediate physical indices are summed, leaving one final
physical index of dimension four. If $b_e,w_e,s_e$ count the MPS bonds, line-MPO bonds, and SWAP
gates on an edge $e$, then
\begin{equation}
    \label{eq:edgeBudget}
    D_e\leq D_1^{b_e}D_U^{w_e}16^{s_e},
    \qquad b_e,w_e,s_e\leq2.
\end{equation}
Taking the maximum over the edges proves the claim.
\end{proof}

\begin{proof}[Proof of Lemma~\ref{lem:pepsError}]
Choose phases with nonnegative overlaps. The squared distance between products of normalized
factors is at most the sum of their squared distances, because
$1-\prod_b a_b\leq\sum_b(1-a_b)$ for $0\leq a_b\leq1$.
Since there are $2m$ chains of
length $m$ at each release, Eq.~\eqref{eq:chainErrorPEPS} therefore gives
\begin{equation}
    \label{eq:resourceErrorAppendix}
    \begin{aligned}
        \bigl\|\ket{\tilde\Phi_{D_1}}-\ket{\Phi}\bigr\|_2^2
        &\leq\frac{8C_0}{D_1}\sum_{m=n/2,n/4,\ldots,2}m^3\\
        &=\frac{8C_0(n^3-8)}{7D_1}.
    \end{aligned}
\end{equation}
At side $n_j$, there are $n_j$ row operations and $n_j$
column operations on the changed species. Lemma~\ref{lem:dModeMPO} and telescoping give
\begin{equation}
    \label{eq:display040}
    \|\tilde W_j-W_j\|_{\mathrm{op}}\leq4\sqrt2\,n_j^2\rho^p.
\end{equation}
Note that all permutations in the exact and approximate constructions are identical and unitary.
Telescoping Eq.~\eqref{eq:completeAnalysis2D} and its approximate version consequently yields
\begin{equation}
    \label{eq:unitaryErrorAppendix}
    \begin{aligned}
        \|\tilde{\mathcal V}_{p}-\mathcal V\|_{\mathrm{op}}
        &\leq4\sqrt2\rho^p\sum_{j=0}^{\ell-2}n_j^2\\
        &=\frac{16\sqrt2}{3}(n^2-4)\rho^p
        <8n^2\rho^p.
    \end{aligned}
\end{equation}
By Eq.~\eqref{eq:resource2D}, the triangle inequality, and unitarity of
$\tilde{\mathcal V}_{p}$, this proves Eq.~\eqref{eq:coverError}.
\end{proof}

% ------------------------------------------------------
\subsection{Transforming the ground state of $\tilde H_2$ to that of $H_2$} \label{sec:appendixH2}

In this Appendix we show that the ground state of $\tilde H_2$ already provides us with that of $H_2$. We use the rotated checkerboard coordinates introduced in Ref. \cite{haegeman2018wavelets}. However, differently from there, we consider a finite periodic lattice. In that case, the transformation to the new coordinates defines a twofold covering, which we describe first.

We rotate the coordinates by $45$ degrees. For that we define a map for the modes $c$ and $d$ which map their coordinates $\boldsymbol r=(x,y)$ to the sites of a new lattice, which we call the physical lattice
\begin{equation}
    \label{eq:coverMap}
    \phi_c(x,y)=(x+y,x-y),\quad
    \phi_d(x,y)=(x+y+1,x-y),
\end{equation}
modulo $n$. The $c$ modes are mapped to the even checkerboard sites while the $d$ modes to the odd ones. In order for $\tilde H_2$ to give the right hopping terms in the new lattice
we redefine $\bar c_{x,y}=(-1)^{x+y}c_{x,y}$, and $\bar d_{x,y}=(-1)^{x+y}d_{x,y}$  which do not influence the PEPS approximation.

Note that each physical site has two preimages differing by $\boldsymbol t=(n/2,n/2)$, i.e., $\phi_{c,d}(\boldsymbol r+\boldsymbol t)=\phi_{c,d}(\boldsymbol r)$. Thus, the new lattice has two modes per site, like the original one. However, we can choose a linear combination thereof where they are decoupled. For that, we define two modes at the physical lattice site $\boldsymbol x$, namely $a_{\boldsymbol x,+},a_{\boldsymbol x,-}$, choosing one representative of each preimage pair:
\begin{equation}
    \label{eq:display041}
    \begin{aligned}
        a_{\phi_c(\boldsymbol r),\pm}
        &=\frac{\bar c_{\boldsymbol r}\pm\bar c_{\boldsymbol r+\boldsymbol t}}{\sqrt2},\\
        a_{\phi_d(\boldsymbol r),\pm}
        &=\frac{\bar d_{\boldsymbol r}\pm\bar d_{\boldsymbol r+\boldsymbol t}}{\sqrt2}.
    \end{aligned}
\end{equation}
One can easily show that the Hamiltonian $\tilde H_2$ does not connect the $\pm$ modes and that gives rise to a tight binding Hamiltonian $H_2$ for each of them, although with different boundary conditions. Indeed, the plus sector is periodic and the minus sector has an antiperiodic twist in each direction. The translationally invariant ground state of $\tilde H_2$ specified above thus factorizes into two Gaussian states
\begin{equation}
    \label{eq:coverFactorization}
    \ket{\Omega_n}=
    \ket{\Omega_n^{\mathrm{per}}}_+\otimes\ket{\Omega_n^{\mathrm{ap,ap}}}_-.
\end{equation}

Now, let us construct the PEPS for the tight binding model. We start out with the original $n\times n$ lattice with a tensor at each site $\mathbf r=(x,y)$:
\begin{equation}
    \label{eq:display042}
    A_{\mathbf r}^{i_c,i_d}(l,r,u,d).
\end{equation}
Each tensor has two physical fermionic indices, $i_c,i_d\in\{0,1\}$, and four virtual indices, each of dimension at most $D$. We now fold the lattice according to the map \eqref{eq:coverMap}, and define a new tensor combining those at position $\boldsymbol r$ and $\boldsymbol r + \boldsymbol t$, i.e. $B_{\mathbf r}=A_{\mathbf r}\otimes A_{\mathbf r+\mathbf t}$.

We now take into account that the two physical modes, $c$ and $d$ belonging to the same original cell are placed at two horizontally neighboring physical sites. Thus, we split the tensor $B_{\mathbf r}$ into two, and put them in a new  physical lattice. Since there are two $d$ modes per site, this will increase the horizontal bond dimension by a factor of 4. The last thing we have to do is to transform the bonds of the original lattice into the ones of the physical lattice. Indeed, those bonds $l,r,u,d$ are now diagonal in the physical lattice. It is simple to rewrite them as horizontal and vertical bonds at the expense of taking the square of the bond dimension. Since the folding already gives another square in the bond indices, the resulting PEPS has bond dimension $D'$ at most $4D^4$.

The final point is to eliminate the sector with antiperiodic boundary conditions, since the PEPS describes both sectors (see Eq.~\eqref{eq:coverFactorization}). Indeed, we have a PEPS $\ket{\hat \Omega}_{D'}$ such that
\begin{equation}
    \label{eq:OmH2approx}
    \| \ket{\hat \Omega}_{D'}-\ket{\Omega_n^{\mathrm{per}}}_+\otimes\ket{\Omega_n^{\mathrm{ap,ap}}}_-\|_2\le \delta
\end{equation}
where $\delta$ is the error in the PEPS approximation of $\ket{\Omega_n}$. We can always expand
\begin{equation}
    \label{eq:display043}
    \ket{\Omega_n^{\mathrm{ap,ap}}}_-= \sum_z b_z \ket{z}.
\end{equation}
where $\ket{z}$ is a Fock (computational) basis. We can also expand
\begin{equation}
    \label{eq:display044}
    \ket{\hat \Omega}_{D'}= \sum_z \ket{\eta_z} \otimes \ket{z}
\end{equation}
If the approximation is exact, $\ket{\eta_z}=b_z\ket{\Omega_n^{\mathrm{per}}}_+$. For an approximation, using Eq.~\eqref{eq:OmH2approx} we have
\begin{equation}
    \label{eq:display045}
    \sum_z \| \ket{\eta_z} - b_z \ket{\Omega_n^{\mathrm{per}}}_+\|_2^2 \le \delta^2
\end{equation}
Since $\sum_z|b_z|^2=1$, there must exist at least one $z$ with $b_z\ne0$ such that
\begin{equation}
    \label{eq:display046}
    \| \ket{\eta_z}/b_z  - \ket{\Omega_n^{\mathrm{per}}}_+ \|_2 \le \delta
\end{equation}
Thus, by finally projecting the $\boldsymbol a_-$ modes in a product state we obtain an unnormalized PEPS approximation of the desired state. Since normalization can only give an extra factor of 2, we obtain a PEPS approximation to the ground state of $H_2$ with bond dimension $D'$ and error $2\delta$.
% ======================================================
\section{Gaussian PEPS approximation of gapped Gaussian Hamiltonians}
\label{sec:appendixGapped}

In this Appendix we prove the results used in Sec.~\ref{sec:gappedGaussianPEPS}. We first bound the error of the repeated one-particle filter and show that an appropriate initial occupation product state always exists. We also give the constant-bond PEPO for one filtering step. Finally, we justify the projection used to extract one copy in the general Gaussian case.

% ------------------------------------------------------
\subsection{Particle number conserving models}
\label{sec:gappedFilterProofs}

We use $J$, $V$, $W$, and $E_S$ as defined in Secs.~\ref{sec:gappedGaussianModels} and~\ref{sec:gappedNumberConserving}. The norms of one-particle matrices in this subsection are operator norms, unless the Frobenius subscript is indicated. We begin by bounding the error of the normalized state $\ket{\Phi_q}$ in Eq.~\eqref{eq:gappedFilteredState}.

\begin{lemma}
    \label{lem:gappedFilterConvergence}
    Suppose that $V^\dagger E_S$ is invertible. For every integer $q\geq0$,
    \begin{equation}
        \label{eq:gappedFilterConvergenceBound}
        \abs{\bra{\Phi}\Phi_q\rangle} \geq \exp\bigg(-\frac{r}{2}\norm{(V^\dagger E_S)^{-1}}^2\bigg(\frac{1 - \Delta}{1 + \Delta}\bigg)^{2q}\bigg)
    \end{equation}
    In particular, an error at most $\epsilon$ is obtained if
    \begin{equation}
        \label{eq:number_K}
        q \geq \frac{\log(\smallnorm{(V^\dagger E_S)^{-1}} + \log(\sqrt{2r/\epsilon})}{\log((1 + \Delta)/(1 - \Delta))}.
    \end{equation}
\end{lemma}
\begin{proof}
We note that 
    \begin{align}\label{eq:J_diag}
        \Id-J = \begin{bmatrix} V &  W \end{bmatrix}\begin{bmatrix}
            \Id-\Lambda_v & 0 \\
            0 & \Id -\Lambda_w
        \end{bmatrix} \begin{bmatrix} V^\dagger \\  W^\dagger \end{bmatrix},
    \end{align}
    where $\Lambda_v$ is a diagonal matrix of eigenvalues of $J$ corresponding to the eigenvectors $v_1, v_2 \dots v_r$ and $\Lambda_w$ is a diagonal matrix of eigenvalues of $J$ corresponding to the eigenvectors $w_1, w_2 \dots w_{N - r}$. Now, using the Slater's determinant formula, we obtain
    \begin{align}
        \abs{\bra{\Phi}\Phi_q\rangle}^2 = \frac{\smallabs{\text{det}(V^\dagger (\Id - J)^q E_S)}^2}{\smallabs{\text{det}(E_S^\dagger (\Id - J)^{2q}E_S}}
    \end{align}
    Using Eq.~\eqref{eq:J_diag}, we obtain
    \begin{align}
        &{\text{det}(E_S^\dagger (\Id - J)^{2q}E_S)}= \nonumber\\
        &\qquad \text{det}\bigg(\sum_{X \in \{V, W\}} E_S^\dagger X(\Id-\Lambda_x)^{2q}X^\dagger E_S \bigg ),
    \end{align}
    and
    \begin{align}
        \text{det}(V^\dagger(\Id - J)^q E_S) &= \text{det}((\Id - \Lambda_v)^q V^\dagger E_S) \nonumber\\
        &=\text{det}(E_S^\dagger V (\Id - \Lambda_v)^q)^*.
    \end{align}
    Finally, using these expressions together with $\text{det}(A^{-1}) = (\text{det}(A))^{-1}$ and $\text{det}(A B C) = \text{det}(A) \text{det}(B) \text{det}(C)$, we then obtain that
    \begin{align}
        \abs{\bra{\Phi}\Phi_q\rangle}^2=(\text{det}(\Id + M^\dagger M))^{-1},
    \end{align}
    where
    \begin{align}
        M = (\Id - \Lambda_w)^q W^\dagger E_S (V^\dagger E_S)^{-1}(\Id-\Lambda_v)^{-q}.
    \end{align}
    From $(\Id-\Lambda_v) \succeq (1 + \Delta)\Id$, $(\Id - \Lambda_w) \preceq (1 - \Delta) \Id$, it follows that
    \begin{align}\label{eq:upper_bound_M}
        \norm{M} \leq  \smallnorm{(V^\dagger E_S)^{-1}} \bigg( \frac{1 - \Delta}{1 + \Delta}\bigg)^q,
    \end{align}
    where we have used that $\norm{W^\dagger E_S} \leq \norm{W}\norm{E_S} \leq 1$. Finally, using the fact that, for a $n\times n$ positive Hermitian matrix $A$, $\abs{\text{det}(A)} \leq \norm{A}^n $, we obtain
    \begin{align}\label{eq:lower_bound_fidelity}
        \abs{\bra{\Phi}\Phi_q\rangle}^2 \geq (1 + \norm{M}^2)^{-r}.
    \end{align}
    To finish the proof, note that if $q$ satisfies Eq.~\eqref{eq:number_K}, then 
    \begin{align}
        \bigg(\frac{1 - \Delta}{1 + \Delta} \bigg)^q &\leq \frac{1}{\smallnorm{(V^\dagger E_S)^{-1}}}\bigg(\frac{\epsilon}{2r}\bigg)^{1/2} \nonumber\\
        &\leq   \frac{1}{\sqrt{r}\smallnorm{(V^\dagger E_S)^{-1}}}(\log((1 - \epsilon)^{-1}))^{1/2},
    \end{align}
    where we have used that, for $x \in (0,1/2 )$, $x\leq 2\log((1 -x)^{-1})$. Therefore, from Eq.~\eqref{eq:upper_bound_M}, it follows that
    \begin{align}
        (1 + \norm{M}^2)^{-r} \geq e^{-r \norm{M}^2} \geq 1 - \epsilon,
    \end{align}
    and thus, from Eq.~\eqref{eq:lower_bound_fidelity}, we obtain that $\abs{\langle \Phi | \Phi_q\rangle}^2 \geq 1 - \epsilon$.
\end{proof}

\noindent The next lemma controls how large $\norm{(V^\dagger E_S)^{-1}}$ must be when the initial product state is chosen appropriately.

\begin{lemma}
    \label{lem:gappedGoodSubset}
    For every $N\times r$ matrix $V$ with $V^\dagger V=1$, there is an $r$-element set $S$ for which $V^\dagger E_S$ is invertible and
    \begin{equation}
        \label{eq:gappedSubsetBound}
        \|(V^\dagger E_S)^{-1}\|_{\mathrm{op}}
        \leq\sqrt{r(N-r+1)}.
    \end{equation}
\end{lemma}
\begin{proof}
We define $Q_S = V^\dagger E_S$: The key idea is to choose $S$ with probability
\begin{equation}
    \label{eq:gappedSubsetDistribution}
    p(S)=\begin{cases}|\det(Q_S)|^2 & \text{ if } Q_S \text{ is invertible}, \\
    0 & \text{ otherwise.}
    \end{cases}.
\end{equation}
$p(S)$ can be viewed as the probability of measuring the computational basis state $S$ in the ground state, and thus $\sum_S p(S) = 1$. Thus
\begin{equation}
    \label{eq:expected_value}
    \begin{aligned}
        \mathbb E_S\|Q_S^{-1}\|_{\mathrm{op}}^2
        &\leq\mathbb E_S\|Q_S^{-1}\|_F^2\\
        &=\sum_{S:p(S)>0}|\det Q_S|^2\|Q_S^{-1}\|_F^2.
    \end{aligned}
\end{equation}
For an invertible matrix, the entries of its inverse are its cofactors divided by its determinant. If $Q_S^{\sim j,\sim i}$ denotes the matrix with row $j$ and column $i$ removed, we obtain
\begin{equation}
    \label{eq:gappedCofactorSum}
    \mathbb E_S\|Q_S^{-1}\|_{\mathrm{op}}^2
    \leq\sum_{|S|=r}\sum_{i,j=1}^{r}
    |\det(Q_S^{\sim j,\sim i})|^2.
\end{equation}
Here we have only added nonnegative terms from singular subsets.

Let $V^{\sim j}$ be $V$ with column $j$ removed. Removing column $i$ from $E_S$ leaves $E_T$ for a set $T$ of $r-1$ sites. For each $T$, there are $N-r+1$ choices of the removed site. Therefore
\begin{align}
    \label{eq:gappedMinorCounting}
    &\sum_{|S|=r}\sum_{i,j=1}^{r}
    |\det(Q_S^{\sim j,\sim i})|^2 =\nonumber\\
    &\qquad \quad (N-r+1)\sum_{j=1}^{r}\sum_{|T|=r-1}
    |\det((V^{\sim j})^\dagger E_T)|^2.
\end{align}
For each $j$, the columns of $V^{\sim j}$ are orthonormal, and therefore
\begin{align}
    \sum_{\abs{T} = r - 1}\smallabs{\text{det}((V^{\sim j})^\dagger E_T)}^2 = 1.
\end{align}
This can be seen from the fact that, from the Slater's formula, $\smallabs{\text{det}((V^{\sim j})^\dagger E_T)}$ is the overlap of the $r-$excitation state specified by the columns of $V^{\sim j}$ with the $r-$excitation computational-basis state specified by $T$. Hence
\begin{equation}
    \label{eq:gappedSubsetExpectation}
    \mathbb E_S\|Q_S^{-1}\|_{\mathrm{op}}^2
    \leq r(N-r+1).
\end{equation}
Finally, note that if the average of $\norm{Q_S^{-1}}_\text{op}^2$ over some probability distribution over $S$ satisfies an upper bound then there is at-least one $S$ which satisfies this bound, proving the lemma. 
\end{proof}

Combining Lemmas~\ref{lem:gappedFilterConvergence} and~\ref{lem:gappedGoodSubset} gives Eq.~\eqref{eq:gappedFilterError}. It remains to show that each filtering step has a constant-bond PEPO representation. This is the same one-step mechanism used for the explicit Chern construction in Appendix~\ref{sec:appendixChern}.

\begin{lemma}
    \label{lem:gappedStepPEPO}
    The operator $Z=e^K$ defined in Eq.~\eqref{eq:gappedFilterGenerator} has an exact fermionic Gaussian PEPO representation of bond dimension $D_0$, where $D_0$ depends only on the spatial dimension, $R$, and the number of modes per site. 
\end{lemma}
\begin{proof}
We set $T=\Id-J$ and introduce an auxiliary mode $g_i$ at the same site as each physical mode $a_i$. With a fixed fermionic ordering, we take
\begin{equation}
    \label{eq:gappedStepChoi}
    \ket{\mathcal I}=\prod_i(1+a_i^\dagger g_i^\dagger)\ket0,
    \qquad \ket{\Psi_Z}=(Z\otimes\Id)\ket{\mathcal 1}.
\end{equation}
Equation~\eqref{eq:gappedFilterAction} gives
\begin{equation}
    \label{eq:gappedStepChoiProduct}
    \ket{\Psi_Z}=\prod_{i,j:T_{ij}\ne0}
    (1+T_{ij}a_i^\dagger g_j^\dagger)\ket0.
\end{equation}
All pair-creation terms are even, commute, and are nilpotent. A factor connecting distinct sites uses one binary fermionic virtual channel; an onsite factor costs no spatial bond. We then route each finite-range connection along a fixed nearest-neighbor path. Since the dimension, range, and number of modes per site are fixed, only a constant number $c$ of these paths can cross any given edge. Fusing their channels gives $D_0\leq2^c$. Note that all Fermionic signs are included in the contractions. Furthermore, the local maps and the Choi state are Gaussian~\cite{bravyi2005}.

The Choi representation gives a PEPO for $Z$ with the same spatial bonds. Contracting $q$ such layers with $\ket{\Phi_S}$ gives a Gaussian PEPS with bond at most $D_0^q$. Normalizing changes one scalar and does not affect this bound.
\end{proof}

% ------------------------------------------------------
\subsection{Projection onto one copy}
\label{sec:gappedProjectionProof}

The onsite change of variables in Eq.~\eqref{eq:change_variables} maps two copies of the general Gaussian Hamiltonian to a particle-number-conserving model. After constructing an approximation and undoing this change, we have a Gaussian PEPS close to a product of the two original ground states. The following observation allows us to retain just one of them. The approximate state itself need not factorize.

\begin{lemma}
    \label{lem:gappedOneCopyProjection}
    Suppose that $\ket{\Xi}$ is a normalized Gaussian PEPS of bond dimension $D$ on two copies of the lattice and
    \begin{equation}
        \label{eq:gappedTwoCopyError}
        \|\ket{\Xi}-\ket{\Omega}\ket{\Omega}\|\leq\delta<1.
    \end{equation}
Then, there is a state that is $\delta$-close to $\ket{\Omega}$ which is also a PEPs of bond dimension $D$.
\end{lemma}
\begin{proof}
Expanding both $\ket{\Xi}$ and $\ket{\Omega}\ket{\Omega}$ in the occupation number basis on the second copy, we can write:
\begin{equation}
    \label{eq:gappedOneCopyExpansions}
    \ket{\Omega}\ket{\Omega} =\sum_X b(X)\ket{\Omega}\ket{X},
    \qquad \ket{\Xi}=\sum_X\ket{\eta_X}\ket X.
\end{equation}
Equation~\eqref{eq:gappedTwoCopyError} implies
\begin{equation}
    \label{eq:gappedOneCopyAverage}
    \sum_X\|\ket{\eta_X}-b(X)\ket{\Omega}\|^2\leq\delta^2,
\end{equation}
which can be rewritten as
\begin{equation}
    \label{eq:gappedOneCopyAverage}
    \sum_X\abs{b(X)}^2 \norm{\frac{\ket{\eta_X}}{\abs{b(X)}}-\ket{\Omega}}^2\leq\delta^2,
\end{equation}
Since $\sum_X|b(X)|^2=1$, there is a $X$ with $b(X)\ne0$ for which
\begin{equation}
    \label{eq:gappedOneCopyRelative}
   \norm{\frac{\ket{\eta_X}}{\abs{b(X)}}-\ket{\Omega}}\leq\delta.
\end{equation}
The state $\eta_X$ is obtained from $\ket{\Xi}$ by performing a computational basis projection on one of the copies, and hence if $\ket{\Xi}$ is a PEPs of bond dimension $D$, so is $\eta_X/\abs{b(X)}$. This proves the lemma statement.
\end{proof}

% ======================================================
\section{Explicit Gaussian PEPS approximation of the Chern band}
\label{sec:appendixChern}
\label{sec:chernPEPS}

In this Appendix we retain the explicit construction for the Chern insulator in Eq.~\eqref{eq:chernHamiltonian}. Unlike the general filtering argument of Sec.~\ref{sec:gappedGaussianPEPS}, it uses the closed expression for the occupied orbitals. We approximate the square root by a polynomial, factor the resulting one-particle map into fixed-range matrices, and represent each induced root map as a Gaussian PEPO. The only approximation is in the occupied orbitals; the subsequent tensor construction is exact.

To proceed, it is first convenient to express the Hamiltonian in momentum space. We group the sites in magnetic cells $r=(x,m)\in\mathcal C$, where $\mathcal C=\{0,\ldots,n-1\}\times\{0,\ldots,n/2-1\}$ and the number of cells is $M=|\mathcal C|=n^2/2$. We write
\begin{equation}
    \label{eq:chernCells}
    \begin{aligned}
        \ket{r,A}&=\ket{x,2m},& a_{r,A}&=a_{x,2m},\\
        \ket{r,B}&=\ket{x,2m+1},& a_{r,B}&=a_{x,2m+1}.
    \end{aligned}
\end{equation}
The momentum-space $ K_{\mathrm C}$ comprises of $(k_x, k_y)$ where
\begin{equation}
    \label{eq:chernMomenta}
    \begin{aligned}
        k_x&=\frac{(2j_x+1)\pi}{n},
        &j_x&=0,\ldots,n-1,\\
        k_y&=\frac{(2j_y+1)\pi}{n/2},
        &j_y&=0,\ldots,n/2-1.
    \end{aligned}
\end{equation}
In Fourier space,
\begin{equation}
    \label{eq:chernFourier}
    \begin{aligned}
        \ket{k,\alpha}
        &=\frac1{\sqrt M}\sum_{r\in\mathcal C}
        e^{ik\cdot r}\ket{r,\alpha},\\
        \hat a_{k,\alpha}^\dagger
        &=\frac1{\sqrt M}\sum_{r\in\mathcal C}
        e^{ik\cdot r}a_{r,\alpha}^\dagger,
        \qquad \alpha=A,B,
    \end{aligned}
\end{equation}
where $k\cdot r=k_xx+k_ym$ and the physical vertical momentum is $k_y/2$, we obtain
\begin{equation}
    \label{eq:chernMomentumHamiltonian}
    H_{\mathrm C}=\sum_{k\in\mathcal K_{\mathrm C}}
    \begin{pmatrix}\hat a_{k,A}^\dagger&
    \hat a_{k,B}^\dagger\end{pmatrix}
    h(k)
    \begin{pmatrix}\hat a_{k,A}\\\hat a_{k,B}\end{pmatrix},
\end{equation}
where
\begin{equation}
    \label{eq:chernBlochHamiltonian}
    h(k)=
    \begin{pmatrix}a(k)&b(k)\\
    \overline{b(k)}&-a(k)\end{pmatrix},
\end{equation}
with
\begin{equation}
    \label{eq:chernBlochEntries}
    \begin{aligned}
        a(k)&=\cos k_x,\\
        b(k)&=\frac{1+e^{-ik_y}+\sin k_x(1-e^{-ik_y})}{2}.
    \end{aligned}
\end{equation}
In particular,
\begin{equation}
    \label{eq:chernDispersion}
        s(k)=a(k)^2+|b(k)|^2 =1+\cos^2 k_x\cos^2(k_y/2),
\end{equation}
where $1\leq s(k)\leq2$ and $h(k)^2=s(k)1$. The spectrum has two bands with energies $\pm\sqrt{s(k)}$.
The occupied orbitals correspond to the negative energies and in momentum space can be written as $u(k)=v(k)/\|v(k)\|$, where
\begin{equation}
    \label{eq:chernExactOrbital}
    v(k)=     \begin{pmatrix}-b(k)\\\sqrt{s(k)}+a(k)\end{pmatrix}.
\end{equation}
The ground state $\ket{\Phi}$ is obtained by filling those orbitals. As in the preceding models, it can be written as
\begin{equation}
    \label{eq:chernExactGroundState}
        \ket{\Phi}=\prod_{k\in\mathcal K_{\mathrm C}}
        \beta_{k}^\dagger\ket0,\qquad
        \beta_{k}^\dagger
        =\sum_{\alpha=A,B}u_\alpha(k)
        \hat a_{k,\alpha}^\dagger.
\end{equation}
We fix the ordering of both the momenta and the cells throughout and omit global phases.

As before, it will also be convenient to express these orbitals in position space. Defining
\begin{equation}
    \label{eq:chernExactEta}
        \eta_r^\dagger=\frac1{\sqrt M}\sum_{k\in\mathcal K_{\mathrm C}}
        e^{-ik\cdot r}\beta_{k}^\dagger,
\end{equation}
we can then write
\begin{equation}
    \label{eq:chernExactPositionState}
    \ket{\Phi}=\prod_{r\in\mathcal C}\eta_r^\dagger\ket0.
\end{equation}

\subsection{Construction}
\label{sec:chernConstruction}

Let $P_p$ be a real polynomial of degree $p\geq1$, whose coefficients are specified below. We will keep the subindices $n,p$ in the state, i.e. $\ket{\Phi_{n,p}}=\ket{\Phi}$. We replace the orbital in Eq.~\eqref{eq:chernExactOrbital} by
$u_p(k)=v_p(k)/\|v_p(k)\|$ with
\begin{equation}
    \label{eq:chernPolynomialOrbital}
    v_p(k)=
    \begin{pmatrix}-b(k)\\P_p(s(k))+a(k)\end{pmatrix}.
\end{equation}
As in Eq.~\eqref{eq:chernExactGroundState}, we define the free fermion state
\begin{equation}
    \label{eq:chernApproximateState}
        \ket{\Phi_{n,p}}=\prod_{k\in\mathcal K_{\mathrm C}} \beta_{k,p}^\dagger\ket0,
        \qquad  \beta_{k,p}^\dagger =\sum_{\alpha=A,B}u_{p,\alpha}(k) \hat a_{k,\alpha}^\dagger.
\end{equation}

In the next subsection, we will show that this state approximates the ground state in Eq.~\eqref{eq:chernExactGroundState} up to an error that decreases exponentially with $p$. We will also show that the state in Eq.~\eqref{eq:chernApproximateState} can be written as a Gaussian PEPS with a bond dimension that grows exponentially with $p$, which will give us the desired result. For that, we need to express this state in a slightly different way in position representation. We define
\begin{equation}
    \label{eq:chernPolynomialEta}
    \begin{aligned}
        \eta_{r,p}^\dagger
        &=\frac1{\sqrt M}\sum_{k\in\mathcal K_{\mathrm C}}
        e^{-ik\cdot r}\|v_p(k)\|_2
        \beta_{k,p}^\dagger\\
        &=\frac1{\sqrt M}\sum_{k,\alpha}
        e^{-ik\cdot r}v_{p,\alpha}(k)
        \hat a_{k,\alpha}^\dagger.
    \end{aligned}
\end{equation}
Note that, unlike the modes in Eq.~\eqref{eq:chernExactEta}, these modes do not obey canonical anticommutation relations. Still, we can write
\begin{equation}
    \label{eq:chernApproximatePositionState}
    \ket{\Phi_{n,p}}\propto \prod_{r\in\mathcal C}\eta_{r,p}^\dagger\ket0.
\end{equation}
Note that the normalization factor does not affect the PEPS bond dimension, so that we will not include it here.

For a one-particle vector $\ket{\phi}=\sum_{r,\alpha}\phi_{r,\alpha}\ket{r,\alpha}$, write $a^\dagger(\phi)=\sum_{r,\alpha}\phi_{r,\alpha}a_{r,\alpha}^\dagger$. We can write the mode operators in Eq.~\eqref{eq:chernPolynomialEta} as
\begin{equation}
    \label{eq:chernEtaPosition}
        \eta_{r,p}^\dagger=a^\dagger(\eta_{r,p})
        =\sum_{r',\alpha}
        (\mathcal T)_{(r',\alpha),(r,B)}a_{r',\alpha}^\dagger.
\end{equation}
with
\begin{equation}
    \label{eq:chernPolynomialFilter}
     \mathcal T=P_p(h^2)-h,
\end{equation}
and $\ket{\eta_{r,p}}=\mathcal T\ket{r,B}$. Here $h$ is the matrix defining the Hamiltonian in Eq.~\eqref{eq:chernHamiltonian}. Indeed, this follows immediately from Eq.~\eqref{eq:chernDispersion}, so that
\begin{equation}
    \label{eq:chernFilterColumn}
    \left[P_p(h(k)^2)-h(k)\right]
    \begin{pmatrix}0\\1\end{pmatrix}
    =v_p(k).
\end{equation}

We now factor the scalar polynomial over the complex numbers:
\begin{equation}
    \label{eq:chernPolynomialRoots}
    r_p(x)=P_p(x^2)-x=c_p\prod_{\nu=1}^{2p}(x-\lambda_\nu),
    \qquad c_p\ne0.
\end{equation}
Note that the roots $\lambda_\nu$ may be complex or repeated and are counted with multiplicity. Setting
\begin{equation}
    \label{eq:chernRootMatrices}
    T_\nu=h-\lambda_\nu1,
    \qquad \mathcal T=c_pT_{2p}\cdots T_1,
\end{equation}
we obtain that the matrix $\mathcal T$ can be written as a product of $2p$ matrices with the same fixed hopping range as $h$.

Now, for each $T_\nu$ matrix, we define the second-quantized operator $Z_\nu$ acting as
\begin{equation}
    \label{eq:chernRootMapDefinition}
    Z_\nu\boldsymbol a^\dagger
    =\boldsymbol a^\dagger T_\nu Z_\nu,
    \qquad Z_\nu\ket0=\ket0.
\end{equation}
These relations uniquely define $Z_\nu$ on ordered products of creation operators, even when $T_\nu$ is singular. Then, one obtains
\begin{equation}
    \label{eq:chernFactorizedState}
    \ket{\Phi_{n,p}}
    =\frac{c_p^M}{\mathcal N_p}
    Z_{2p}\cdots Z_1\ket{R_B}.
\end{equation}
where $\mathcal N_p=\prod_{k\in\mathcal K_{\mathrm C}}\|v_p(k)\|_2>0$ and
\begin{equation}
    \label{eq:chernReferenceState}
    \ket{R_B}=\prod_{r\in\mathcal C}a_{r,B}^\dagger\ket0,
\end{equation}
in which all $B$ sites are occupied and all $A$ sites are empty. This is the desired representation as a product of root maps, induced by fixed-range one-particle matrices, acting on a product state.

% ------------------------------------------------------

\subsubsection{Polynomial approximation of the square root}
\label{sec:chernPolynomial}

We use the standard binomial expansion of $(1+z)^{1/2}$ for $|z|<1$~\cite{dlmfBinomial}, truncated at order $p$,
\begin{equation}
    \label{eq:chernSquareRootPolynomial}
    P_p(s)=\sqrt2\sum_{j=0}^{p}
    \binom{1/2}{j}\left(\frac{s}{2}-1\right)^j,
\end{equation}
where $s\in[1,2]$ and
\begin{equation}
    \label{eq:chernBinomialCoefficient}
    \binom{1/2}{0}=1,\qquad
    \binom{1/2}{j}=\frac{\prod_{a=0}^{j-1}(1/2-a)}{j!},
    \quad j\geq1.
\end{equation}
For $s\in[1,2]$, all nonconstant terms of the expansion are nonpositive and $|\binom{1/2}{j}|\leq1$, so the omitted tail gives
\begin{equation}
    \label{eq:chernPolynomialError}
    \delta_p=\sqrt2\,2^{-p}= \sqrt2\sum_{j=p+1}^{\infty}2^{-j}\geq P_p(s)-\sqrt{s}
    \geq 0.
\end{equation}

% ------------------------------------------------------
\subsubsection{Error in the normalized ground state}
\label{sec:chernGroundStateError}

Here we prove the error estimate used in Sec.~\ref{sec:chernBondAndError}.

\begin{lemma}
    \label{lem:chernStateError}
    Let $n\geq4$ be even and choose $P_p$ as in
    Eq.~\eqref{eq:chernSquareRootPolynomial}. For the states in
    Eqs.~\eqref{eq:chernExactGroundState} and
    \eqref{eq:chernApproximateState}
    \begin{equation}
        \label{eq:chernStateErrorLemma}
        \bigl\|\ket{\Phi_{n,p}}-\ket{\Phi_n}\bigr\|_2
        \leq n^2 2^{-p}.
    \end{equation}
\end{lemma}
\begin{proof}
For each  momentum, we write
\begin{equation}
    \label{eq:chernErrorColumns}
    v=\begin{pmatrix}-b\\\sqrt{s}+a\end{pmatrix},
    \qquad
    v_p=\begin{pmatrix}-b\\P_p(s)+a\end{pmatrix},
\end{equation}
where $a=\cos k_x$,
$b=[1+e^{-ik_y}+\sin k_x(1-e^{-ik_y})]/2$, and
$s=a^2+|b|^2$. On the antiperiodic grid of Eq.~\eqref{eq:chernMomenta},
$|\sin k_x|\geq\sin(\pi/n)\geq2/n$. Therefore
\begin{equation}
    \label{eq:chernColumnBounds}
    \|v\|\geq2/n,
    \qquad \|v_p-v\|\leq\delta_p.
\end{equation}
For nonzero vectors $v,w$, we can use
\begin{equation}
    \label{eq:chernNormalizationInequality}
    \left\|\frac{w}{\|w\|}-\frac{v}{\|v\|}\right\|
    \leq\frac{2\|w-v\|}{\|v\|}.
\end{equation}
Thus
\begin{equation}
    \label{eq:chernNormalizedOrbitalError}
    \|u_p(k)-u(k)\|_2\leq n\delta_p.
\end{equation}

The orbital overlap is
\begin{equation}
    \label{eq:chernOrbitalOverlap}
    r_{k}=u(k)^\dagger u_p(k)
    =\frac{|b|^2+(\sqrt{s}+a)(P_p(s)+a)}
    {\|v\|\|v_p\|}.
\end{equation}
Here, $0<r_{k}\leq1$ since $s>a^2$ on the  momentum grid and $P_p(s)\geq\sqrt{s}$. Since different momentum sectors are orthogonal, the overlap of the Slater determinants is
\begin{equation}
    \label{eq:chernSlaterOverlap}
    \langle\Phi_n|\Phi_{n,p}\rangle
    =\prod_{k\in\mathcal K_{\mathrm C}}r_{k}.
\end{equation}
Using $1-\prod_jr_j\leq\sum_j(1-r_j)$ for $0\leq r_j\leq1$,
we obtain
\begin{equation}
    \label{eq:chernStateErrorProof}
    \begin{aligned}
        \bigl\|\ket{\Phi_{n,p}}-\ket{\Phi_n}\bigr\|_2^2
        &\leq2\sum_{k}(1-r_{k})
        =\sum_{k}\|u_p(k)-u(k)\|_2^2\\
        &\leq M n^2\delta_p^2=n^4 2^{-2p}.
    \end{aligned}
\end{equation}
which proves the claim.
\end{proof}

% ------------------------------------------------------
\subsubsection{Gaussian PEPO representation of the root maps}
\label{sec:chernChoi}

Here we show that $Z_\nu$ in Eq.~\eqref{eq:chernRootMapDefinition} can be written exactly as a PEPO with constant bond dimension. As in previous cases, we use the fermionic Choi construction~\cite{bravyi2005}.

\begin{lemma}
    \label{lem:chernRootPEPO}
    Each operator $Z_\nu$ defined in
    Eq.~\eqref{eq:chernRootMapDefinition} has an exact fermionic Gaussian
    PEPO representation on the original square lattice with
    \begin{equation}
        \label{eq:chernRootPEPOBound}
        D_{Z_\nu}\leq64.
    \end{equation}
\end{lemma}
\begin{proof}
We define $t_{ij}=(T_\nu)_{ij}$. For $i\ne j$, it has the hopping pattern of $h$, while $t_{ii}=-\lambda_\nu$. We also introduce an auxiliary fermionic mode operator $g_i$ at each site, with standard canonical anticommutation relations. We take the
unnormalized maximally entangled state
\begin{equation}
    \label{eq:chernMaximallyEntangledState}
    \ket{\mathcal I}
    =\prod_i(1+a_i^\dagger g_i^\dagger)\ket0.
\end{equation}
The Choi vector is
\begin{equation}
    \label{eq:chernChoiState}
    \ket{\Psi_\nu}=(Z_\nu\otimes1)\ket{\mathcal I}.
\end{equation}
Using Eq.~\eqref{eq:chernRootMapDefinition}, with $Z_\nu$ acting only on the physical modes, we can write
\begin{equation}
    \label{eq:chernChoiFactorization}
        \ket{\Psi_\nu}
        =\prod_{i,j:t_{ij}\ne0}
        (1+t_{ij}a_i^\dagger g_j^\dagger)\ket0.
\end{equation}
The even pair-creation terms commute and are nilpotent, which gives the product in Eq.~\eqref{eq:chernChoiFactorization}. A PEPS for this vector yields a PEPO for $Z_\nu$ with the same bond dimensions; the construction uses Gaussian local maps.

For $i\ne j$, each factor $1+t_{ij}a_i^\dagger g_j^\dagger$ uses one binary fermionic virtual channel, of dimension two. Terms with $i=j$ are onsite and cost no spatial bond. Route axial terms directly and each directed diagonal term from $j$ to $i$ first horizontally and then vertically. At most two axial channels and four directed diagonal channels cross any nearest-neighbor edge. Fusing them gives $D_{Z_\nu}\leq2^{2+4}=64$. Fermionic wire permutations and antiperiodic seam signs change coefficients, not bond dimensions.
\end{proof}

% ======================================================================

\subsection{Bond-dimension and error}
\label{sec:chernBondAndError}

In Sec.~\ref{sec:chernChoi} we showed that each $Z_\nu$ admits an exact Gaussian PEPO with bond dimension at most $D_0=64$, independently of $n$, $p$, and $\lambda_\nu$. Thus, their product will yield a Gaussian PEPS with
\begin{equation}
    \label{eq:chernPEPSBond}
    D\leq D_0^{2p}=2^{12p}.
\end{equation}

For the polynomial chosen in Sec.~\ref{sec:chernPolynomial} above, Lemma~\ref{lem:chernStateError} gives \begin{equation}
    \label{eq:chernPEPSError}
    \bigl\|\ket{\Phi_{n,p}}-\ket{\Phi_n}\bigr\|_2
    \leq n^2 2^{-p}.
\end{equation}
This is a bound on the normalized entire wavefunction. Together, these estimates give the following result. As before, the bounds are conservative and are intended to establish polynomial scaling. Unlike the critical-state approximation, the PEPS constructed here is Gaussian.

\begin{theorem}
    \label{thm:chernPEPS}
    Let $n\geq4$ be even and $0<\epsilon<1$. The ground state
    $\ket{\Phi_n}$ of Eq.~\eqref{eq:chernHamiltonian}, with antiperiodic
    boundary conditions in both directions, has a normalized one-mode
    fermionic Gaussian PEPS approximation $\ket{\Phi_{n,p}}$ satisfying
    \begin{equation}
        \label{eq:chernEpsilonApproximation}
        \bigl\|\ket{\Phi_{n,p}}-\ket{\Phi_n}\bigr\|_2\leq\epsilon
    \end{equation}
    with bond dimension $D=\operatorname{poly}(n,1/\epsilon)$.
\end{theorem}
\begin{proof}
Choose
\begin{equation}
    \label{eq:chernApproximationDegree}
    p=\left\lceil\frac{\log(n^2/\epsilon)}{\log2}\right\rceil.
\end{equation}
Lemma~\ref{lem:chernStateError} gives
Eq.~\eqref{eq:chernEpsilonApproximation}, while
Lemma~\ref{lem:chernRootPEPO} and Eq.~\eqref{eq:chernFactorizedState}
give Eq.~\eqref{eq:chernPEPSBond}. Since $2^p\leq2n^2/\epsilon$,
\begin{equation}
    \label{eq:chernExplicitBond}
    D\leq4096\,n^{24}\epsilon^{-12},
\end{equation}
which proves the claim.
\end{proof}

% ======================================================
% ======================================================
\section{Proofs for the Gutzwiller projection}
\label{sec:GutzwillerLemmas}

In this Appendix we prove the stability estimate used in Sec.~\ref{sec:GutzwillerProjected}. We keep a fixed number $m$ of copies and work in the physical occupation basis. The main point is to bound the change in the normalized projected state relative to its own unnormalized norm, so that a small projection probability does not enter the result.

In the $r$-particle sector of $N$ modes, fix an ordering of the physical operators $a_x$. For an $r$-element set $X$, write
\begin{equation}
    \label{eq:decomposition_Phi}
    \ket{\Phi}=\sum_{|X|=r}\phi(X)\ket X,
    \qquad \ket X=\prod_{x\in X}a_x^\dagger\ket0.
\end{equation}
We use the coefficient norms
\begin{equation}
    \label{eq:gutzCoefficientNorm}
    \|\Phi\|_p=\left(\sum_X|\phi(X)|^p\right)^{1/p},
    \qquad p\geq1,
\end{equation}
and omit the subscript when $p=2$. These norms refer to this fixed physical basis, not to the paired modes of Eq.~\eqref{eq:tildebetai}. For two coefficient vectors, denote their componentwise product by
\begin{equation}
    \label{eq:gutzComponentProduct}
    \ket{\Phi\odot\widetilde\Phi}
    =\sum_X\phi(X)\widetilde\phi(X)\ket X.
\end{equation}
We also use $\ket{\Phi^m}$ for the product of $m$ identical coefficient vectors; this is not a tensor power. The normalized projected state in Eq.~\eqref{eq:gutzPowerState} is
\begin{equation}
    \label{eq:gutzPowerNormalization}
    \ket{\Psi}=\frac{\ket{\Phi^m}}{\|\Phi^m\|},
    \qquad \|\Phi^m\|=\|\Phi\|_{2m}^{m}.
\end{equation}
For suitable filled $C=1$ bands this is the usual lattice Laughlin-type power construction. The estimates below do not require a topological interpretation of the input or output state.

For an operator $\mathcal O$, define the induced norm
\begin{equation}
    \label{eq:inducedPnorm}
    \|\mathcal O\|_{p\to p}
    =\sup_{\Phi\ne0}\frac{\|\mathcal O\ket{\Phi}\|_p}{\|\Phi\|_p}.
\end{equation}
It is submultiplicative. We consider the path
\begin{equation}
    \label{eq:gutzGaussianPath}
    \ket{\Phi(\theta)}=e^{\theta\hat A}\ket{\Phi},
    \qquad 0\leq\theta\leq1,
\end{equation}
where, in the original physical modes,
\begin{equation}
    \label{eq:decomp_A_hat}
    \hat A=\sum_{x,y=1}^{N}C_{xy}a_x^\dagger a_y,
    \qquad C^\dagger=-C.
\end{equation}
In the application $C=UAU^\dagger$ is traceless, but the following estimates do not need that extra property. Let $\ket{\Psi(\theta)}$ be the normalized componentwise $m$th power of $\ket{\Phi(\theta)}$. Our goal is to bound the distance between $\ket{\Psi(1)}=\ket{\tilde\Psi}$ and $\ket{\Psi(0)}=\ket{\Psi}$.

\begin{lemma}
    \label{lem:ErrorPsi1}
    For $0\leq\theta\leq1$,
    \begin{equation}
        \label{eq:lemmaerrorPsi}
        \|\ket{\Psi(\theta)}-\ket{\Psi(0)}\|
        \leq m\int_0^\theta
        \frac{\|\frac{d}{d\theta'}\Phi(\theta')\|_{2m}}
        {\|\Phi(\theta')\|_{2m}}\,d\theta'.
    \end{equation}
\end{lemma}
\begin{proof}
The triangle inequality gives
\begin{equation}
    \label{eq:integral_triangle_inequality}
    \|\ket{\Psi(\theta)}-\ket{\Psi(0)}\|
    \leq\int_0^\theta\left\|\frac{d}{d\theta'}\ket{\Psi(\theta')}\right\|d\theta'.
\end{equation}
To control the derivative, put $w(\theta)=\Phi(\theta)^m$. Differentiating $\Psi=w/\|w\|$ gives
\begin{equation}
    \label{eq:upper_bound_normalized_derivative}
    \left\|\frac{d\Psi}{d\theta}\right\|^2
    =\frac{\|w'\|^2}{\|w\|^2}
    -\frac{(\operatorname{Re}\langle w,w'\rangle)^2}{\|w\|^4}
    \leq\frac{\|w'\|^2}{\|w\|^2}.
\end{equation}
On the other hand, the componentwise product rule and H\"older's inequality imply
\begin{equation}
    \label{eq:derivative_power_Phi}
    \begin{aligned}
        \|w'(\theta)\|
        &=m\|\Phi(\theta)^{m-1}\odot\Phi'(\theta)\|_2\\
        &\leq m\|\Phi(\theta)^{m-1}\|_{2m/(m-1)}
        \|\Phi'(\theta)\|_{2m}\\
        &=m\|\Phi(\theta)\|_{2m}^{m-1}
        \|\Phi'(\theta)\|_{2m}.
    \end{aligned}
\end{equation}
Since $\|w(\theta)\|=\|\Phi(\theta)\|_{2m}^{m}$, division cancels all but one power of the coefficient norm. Substitution into Eq.~\eqref{eq:integral_triangle_inequality} proves the claim. The denominator is nonzero because $\ket{\Phi(\theta)}$ is normalized.
\end{proof}

It will also be useful to bound the induced norms of elementary hopping operators. Their simple action in the occupation basis is the reason no exponential dimension factor appears.

\begin{lemma}
    \label{lem:gutzHoppingNorms}
    Let $a_x$ be canonical fermionic modes. For $p\geq1$, $x\ne y$, and any real $\varphi$,
    \begin{equation}
        \label{eq:gutzHoppingNorms}
        \|a_x^\dagger a_x\|_{p\to p}\leq1,
        \qquad
        \|e^{i\varphi}a_x^\dagger a_y+\mathrm{H.c.}\|_{p\to p}\leq1.
    \end{equation}
    The same bound holds for a minus sign before the Hermitian conjugate.
\end{lemma}
\begin{proof}
The number operator selects the configurations containing $x$:
\begin{equation}
    \label{eq:gutzNumberAction}
    a_x^\dagger a_x\ket X=\mathcal I_X(x)\ket X.
\end{equation}
Thus it cannot increase any coefficient $p$-norm. For $x\ne y$, a hopping pair exchanges the occupations $10$ and $01$, with coefficients of modulus one, and annihilates $00$ and $11$. More explicitly, let $\pi_{xy}$ be the permutation of configurations exchanging the occupations at $x,y$. The action is
\begin{equation}
    \label{eq:gutzSwapAction}
    (e^{i\varphi}a_x^\dagger a_y+\mathrm{H.c.})\ket X
    =f_X\ket{\pi_{xy}(X)},\qquad |f_X|\leq1.
\end{equation}
The fermionic reordering sign is included in $f_X$. Since $\pi_{xy}$ is a bijection,
\begin{equation}
    \label{eq:gutzSwapNormProof}
    \|(e^{i\varphi}a_x^\dagger a_y+\mathrm{H.c.})\Phi\|_p^p
    =\sum_X|f_X|^p|\phi(X)|^p\leq\|\Phi\|_p^p.
\end{equation}
The minus sign only changes the phases of the two nonzero matrix elements, so the proof is identical.
\end{proof}

With these two lemmas, we can bound the projected-state error directly in terms of the one-particle generator.

\begin{lemma}
    \label{lem:MainLemmaGutzwiller}
    For the states and generator above,
    \begin{equation}
        \label{eq:gutzPowerStabilityLemma}
        \|\ket{\tilde\Psi}-\ket{\Psi}\|
        \leq mN\|C\|_F.
    \end{equation}
\end{lemma}
\begin{proof}
Since $\Phi'(\theta)=\hat A\Phi(\theta)$,
\begin{equation}
    \label{eq:gutzRelativeDerivative}
    \frac{\|\Phi'(\theta)\|_{2m}}{\|\Phi(\theta)\|_{2m}}
    \leq\|\hat A\|_{2m\to2m}.
\end{equation}
Lemma~\ref{lem:ErrorPsi1} therefore gives
\begin{equation}
    \label{eq:twolemmas}
    \|\ket{\tilde\Psi}-\ket{\Psi}\|
    \leq m\|\hat A\|_{2m\to2m}.
\end{equation}
Using $C^\dagger=-C$, write
\begin{equation}
    \label{eq:gutzGeneratorPairs}
    \hat A=\sum_x C_{xx}a_x^\dagger a_x
    +\sum_{x<y}(C_{xy}a_x^\dagger a_y-\mathrm{H.c.}).
\end{equation}
The triangle inequality and Lemma~\ref{lem:gutzHoppingNorms} imply
\begin{equation}
    \label{eq:gutzGeneratorInducedBound}
    \|\hat A\|_{2m\to2m}
    \leq\sum_x|C_{xx}|+\sum_{x<y}|C_{xy}|
    \leq\sum_{x,y}|C_{xy}|.
\end{equation}
Finally, Cauchy--Schwarz on the $N^2$ entries gives
\begin{equation}
    \label{eq:gutzGeneratorEntryBound}
    \sum_{x,y}|C_{xy}|\leq N\|C\|_F.
\end{equation}
Combining the last three equations proves the lemma. In Sec.~\ref{sec:GutzwillerProjected}, $\|C\|_F=\|A\|_F$ because the two matrices are related by a unitary change of one-particle basis. No such basis invariance is assumed for the coefficient $p$-norms.
\end{proof}

\FloatBarrier
\renewcommand{\bibsection}{

\end{document}